\documentclass[onecolumn,amsmath,amssymb,titlepage]{revtex4-1}

\usepackage{bbold}
\usepackage{mathptmx}
\usepackage{subfigure}
\usepackage{psfrag,graphicx}
\usepackage{dcolumn}
\usepackage{amsmath,amssymb}
\usepackage{bm}
\usepackage{color}
\usepackage{latexsym}
\usepackage{epstopdf}
\usepackage{color}
\usepackage[english]{babel}
\usepackage{latexsym}
\usepackage{psfrag,graphicx}
\usepackage{subfigure}
\usepackage{amsmath}
\usepackage{amssymb}
\usepackage{amsfonts}
\usepackage{bm}
\usepackage{natbib}
\usepackage{epstopdf}
\usepackage{appendix}
\usepackage{rotating}

\definecolor{mygrey}{gray}{0.35}
\definecolor{myblue}{rgb}{0.2,0.2,0.8}
\definecolor{myzard}{cmyk}{0,0,0.05,0}
\definecolor{mywhite}{rgb}{1,1,1}
\definecolor{mywhite}{rgb}{1,1,1}
\definecolor{myred}{rgb}{1,0.,0.3}

\usepackage[colorlinks=true,citecolor=myblue,linkcolor=myred]{hyperref}

\def\ba{\begin{align}}
\def\enda{\end{align}}
\def\bi{\begin{itemize}}
\def\ei{\end{itemize}}

\def\be{\begin{equation}}
\def\ee{\end{equation}}
\def\bea{\begin{eqnarray}}
\def\eea{\end{eqnarray}}
\def\bse{\begin{subequations}}
\def\ese{\end{subequations}}

\newcommand{\ket}[1]{|{#1}\rangle}                       
\newcommand{\bra}[1]{\langle {#1}|}                      

\newcommand{\Ignore}[1]{ }

\DeclareMathOperator{\sech}{sech}

\begin{document}

\title{Dynamics of quantum discord of two coupled spin-1/2's subjected to time-dependent magnetic fields}

\author{Roberto Grimaudo}
 \affiliation{ Dipartimento di Fisica e Chimica dell'Universit\`a di Palermo, Via Archirafi, 36, I-90123 Palermo, Italy}
\affiliation{ INFN, Sezione Catania, \textit{I-95123} Catania, Italy}

\author{Tatiana Mihaescu}
\affiliation{Faculty of Physics, University of Bucharest, Bucharest-Magurele, Romania}
\affiliation{National Institute of Physics and Nuclear Engineering, Bucharest-Magurele, Romania}

\author{Iulia Ghiu}
\affiliation{ University of Bucharest, Faculty of Physics, Centre for Advanced Quantum Physics, PO Box MG-11, R-077125, Bucharest-Magurele, Romania}

\author{Aurelian Isar}
\affiliation{Faculty of Physics, University of Bucharest, Bucharest-Magurele, Romania}
\affiliation{National Institute of Physics and Nuclear Engineering, Bucharest-Magurele, Romania}

\author{Antonino Messina}
\affiliation{ INFN, Sezione Catania, \textit{I-95123} Catania, Italy}
\affiliation{ Dipartimento di Matematica ed Informatica dell'Universit\`a di Palermo, Via Archirafi, 34, I-90123 Palermo, Italy}

\begin{abstract}
We describe the dynamics of quantum discord of two interacting spin-1/2's subjected to controllable time-dependent magnetic fields.
The exact time evolution of discord is given for various input mixed states consisting of classical mixtures of two Bell states.
The quantum discord manifests a complex oscillatory behaviour in time and is compared with that of quantum entanglement, measured by concurrence.
The interplay of the action of the time-dependent magnetic fields and the spin-coupling mechanism in the occurrence and evolution of quantum correlations is examined in detail.
\end{abstract}

\maketitle

\section{Introduction}
\label{intro}

Quantum correlations became in the last two decades a field of large interest, due to their crucial role played in the quantum information science \cite{nie,ade1}. Much effort and work are presently devoted to characterize and quantify the quantum correlations, like entanglement, steering and discord existing in multipartite quantum states \cite{hor1,giro,ber,hu}. These quantum correlations are considered useful physical quantum resources with promising applications in quantum information processing and transmission tasks and protocols.
It is well known that quantum entanglement does not describe all the properties of non-classical nature of the quantum correlations. In this respect, quantum discord has been proposed as a measure of quantum correlations, beyond entanglement \cite{zur,ollivier}, which can also exist in separable mixed states. The physical understanding of quantum discord constantly advanced and several operational interpretations to discord have been proposed \cite{ani,cav}.

It is nowadays possible to realize physical scenarios where quantum coherence turns out to be robustly protected against detrimental classical and quantum  uncontrollable sources.
This circumstance has  spurred a growing interest toward the quantum dynamics of closed bipartite physical systems subjected to controllable  time dependent  external classical fields.
When the corresponding Hamiltonian model is both non trivial and exactly solvable, one might indeed  undertake  a systematic, hopefully exact, study of the unitary time evolution of correlations get established in the closed system, not traceable back to classical physics \cite{grimess,GMIV,GBNM,GLSM,GVM1,GVM2}.
To follow and to interpret, for example, the appearance, time variation and  death at finite time instants of entanglement and quantum discord is of relevance from both a theoretical and applicative point of view.
On the one hand, such a  knowledge may significantly contribute to highlight the meaning of crucial concepts like non locality and dechoerence and to capture their connection with properties experimentally exhibited by the system.
On the other hand, due to such an interpretative potentiality, it provides the key to clarify the role of quantum correlations as resources for quantum technologies.

In the previous paper \cite{grimess} the authors studied the behaviour of two coupled spin-1/2's interacting with a time-dependent magnetic field. An exact time evolution of the compound system has been obtained and several physically relevant quantities have been evaluated, in particular the concurrence as a measure of the intensity of entanglement. In Refs. \cite{mar,adisc} the evolution of quantum discord and entanglement of a two-mode Gaussian state, namely the squeezed thermal state, in contact with local thermal reservoirs was investigated. It was shown that the discord can increase in time above its initial value in a special situation.
In this work we study the evolution of discord of the physical system described by the model introduced and analyzed in Ref. \cite{grimess}. It is worth to be mentioned that due to the $C_2$-symmetry with respect to the quantization axis $z,$ possessed by this model, it is possible to solve exactly the dynamics of the considered system, by reducing the problem to two independent problems of single spin-1/2.

We will show that, thanks to peculiar symmetries of the Hamiltonian model, the states which describe the evolution of the considered system keeps the form of X-states. Consequently in order to determine the behaviour of the quantum discord we make use of the algorithm known for its evaluation in the case of X-states \cite{ali,li}.
Our analysis discloses the existence of sudden death and birth of both concurrence and quantum discord.
In addition, we highlight the interplay between externally acting time-dependent magnetic fields and the internal coupling mechanism between the two spins in the emergence and evolution of quantum discord. This novelty provided by the behaviour of the mentioned nonclassical correlations - entanglement and discord - could enable a better understanding and description of the intrinsic quantum nature of the considered system.
 
The paper is organized as follows.
In Sec. 2 we introduce the quantum discord for X-states, which is a measure of all quantum correlations in the bipartite state, including entanglement.
In Sec. 3 we introduce the time-dependent models of two interacting spin-1/2's \cite{grimess}.
In Sec. 4 we exactly describe the evolution of quantum discord and concurrence, considering various input mixed states consisting of classical mixtures of two Bell states.
It is emphasized the role played by the magnetic fields in determining the occurrence of quantum correlations between the two spins.
The quantum discord manifests a complex oscillatory behaviour in time and it is compared with that of quantum entanglement, measured by concurrence.
It is shown that for pure initial states the two types of correlations manifest a similar behaviour, however for an initial mixed state they exhibit remarkable differences.
Final remarks and conclusions are given in Sec. 5.

\section{Quantum discord of X-states}
\label{sec:2}

The main feature of the quantum world, discriminating it from the classical one, is the possibility of representing a pure state as superposition of pure states.
Indeed, while a quantum state of a bipartite system is not necessarily writeable as a tensor product of two independent pure states of the two subsystems, a pure state of a classical bipartite system turns out to be always factorizable since the superposition principle does not hold in this context \cite{ade1}.
This crucial difference leads to the following remarkable physical consequences.
1) Contrary to what happens in classical Physics, the knowledge that a non-factorizable state of a quantum bipartite system is pure does not lead to pure states of the two subsystems.
2) Non-factorizable states of a quantum bipartite system lead to non-locality effects christened by Schr\"odinger as entanglement.

Quantum entanglement represents one of the most important resources
in quantum information \cite{nie,hor1}. At the same time, there exist quantum correlations, different from entanglement, with potential applications in quantum information tasks, for instance quantum nonlocality without entanglement \cite{hor1,bennett,niset}.
Likewise, it was shown that there exist separable states which can produce a speeding of some protocols, in comparison to the classical states \cite{datta,datta1,datta2,lanyon}.

Such a kind of nonlocal correlation, introduced by
Ollivier and Zurek \cite{zur,ollivier}, is quantum discord, that received a lot of attention in the recent years \cite{ollivier,ali,luo1,luo,sarandy,werlang,Maziero,fanchini,ferraro,modi,Mazzola,lang,dakic}.
Quantum discord {{is defined as the difference
between two different quantum analogues of classically equivalent expressions of
the quantum mutual information, which is a measure of all correlations in a quantum state.
In a bipartite state discord measures the total quantum correlations, without restricting to entanglement.
Discord coincides with the entropy of entanglement for pure entangled states. Some mixed separable states can have non-zero discord, so that it is considered to represent a  characteristic of the quantumness of such separable states.

The measure of the total correlations in a bipartite system $AB$ is given by the quantum mutual information \cite{gro}
\begin{eqnarray}
\mathcal{I}(\rho^{AB})=S(\rho^A)+S(\rho^B)-S(\rho^{AB}).
\end{eqnarray}
where $S(\rho)=-{\rm Tr}(\rho\log_2\rho)$ is the von Neuman entropy.
$\rho^{AB}$ represents the density operator of the compound system $A+B$, whilst $\rho^{A(B)}={\rm Tr}_{B(A)}(\rho^{AB})$ is the reduced density matrix of the subsystem $A$ ($B$).
Quantum discord was determined \cite{ollivier,luo} by using a measurement-based conditional density operator in order to generalize the classical mutual information.
The considered von Neumann-type measurement consists of one-dimensional local projectors summing to identity.
The quantum mutual information corresponding to the quantum conditional entropy associated to a measurement
\begin{eqnarray}
S(\rho|B_k)=\sum_{k}p_kS(\rho_k),
\end{eqnarray}
is given by \cite{ollivier,hen}
\begin{eqnarray}
\mathcal{I}(\rho|B_k )=S(\rho^A)-S(\rho|B_k).
\end{eqnarray}
Here ${B_k}$ is the set of the projectors which perform the measurement on the subsystem $B$ and $p_k={\rm Tr}(I\otimes B_k)\rho (I\otimes B_k)$ is the measurement probability for the $k$th projector.
We may write, indeed, the conditional density operator $\rho_k$, denoting the reduced density operator of subsystem $A$ after the local measurements and which is associated with the measurement outcome $k$, in the following form ($I$ denotes the identity operator on the subsystem $A$):
\begin{eqnarray}
\rho_k=\frac{1}{p_k}(I\otimes B_k)\rho (I\otimes B_k).
\end{eqnarray}

Quantum discord is interpreted as a measure of quantum correlations since it is defined by the difference between the mutual information $\mathcal{I}(\rho)$ and the classical correlations $\mathcal{C}(\rho)$
\begin{eqnarray}
{D}(\rho)=\mathcal{I}(\rho)-\mathcal{C}(\rho).
\end{eqnarray}
The measure of bipartite classical correlations $\mathcal{C}(\rho)=\sup_{B_k}\mathcal{I}(\rho|B_k)$ (sup is taken over all possible von Neumann local measurements ${B_k}$) represents the quantum mutual information induced by measurement.

When a bipartite system is in a pure state, entanglement and quantum discord give the same information on quantum correlations, while in a mixed state there might be quantum correlations - discord, even if the two subsystems are not entangled.
Quantum discord has the peculiarity to be strictly related to the subsystem under measurement to investigate the existence of quantum correlations.
This means that for a bipartite system composed by two subsystems $A$ and $B$, we speak of quantum discord with respect to the subsystem $A$, ${D}_A$, and $B$, ${D}_B$.
It is useful to remind that quantum discord is zero for a general state of a bipartite system, ${D}_B(\rho_{AB})=0$, if and only if the state can be written as
\begin{equation}
\rho_{AB}=\sum_i p_i \rho_A^i  \otimes \ket{i}\bra{i}_B, \qquad \sum_i p_i=1, \quad p_i\geq 0,
\end{equation}
that is if there exist an orthonormal basis for the subsystem with respect to which we calculate the quantum discord ($B$ in this case) such that its state results diagonal.

The difficulty in calculating quantum discord consists in the complexity of the maximization procedure for computing the classical correlations, due to the fact that maximization has to be performed over all possible von Neumann measurements on party $B$.
Analytical expressions for classical correlations and quantum discord are known for two-qubit Bell diagonal state and for some kinds of two-qubit $X$ states \cite{ali,luo}.

A generic X-state of a system of two spin-1/2, $A$ and $B$, may be cast in the following form
\begin{equation}\label{X-state}
\rho_X=
\left(
\begin{array}{cccc}
 \rho_{11} & 0 & 0 & \rho_{14} \\
 0 & \rho_{22} & \rho_{23} & 0 \\
 0 & \rho_{32} & \rho_{33} & 0 \\
 \rho_{41} & 0 & 0 & \rho_{44} \\
\end{array}
\right).
\end{equation}
The unit trace and positivity conditions read $\sum_{i=1}^4 \rho_{ii}=1$, $\rho_{11}\rho_{44}\geqslant|\rho_{14}|^2$ and $\rho_{22}\rho_{33}\geqslant|\rho_{23}|^2$, assuming in general $\rho_{14}=|\rho_{14}|e^{i\phi_{14}}$ and $\rho_{23}=|\rho_{23}|e^{i\phi_{23}}$.
It is easy to show \cite{cil,mod,huang,sab,yur,cele} that with the help of the local unitary transformation $U_{A}\otimes U_{B}$, such that
\begin{equation}\label{Loc Unit Transf}
U_A=
\begin{pmatrix}
e^{-i(2\phi_{23}+\phi_{14})/4} & 0 \\
0 & e^{i\phi_{14}/4}
\end{pmatrix}, \quad
U_B=
\begin{pmatrix}
e^{i(2\phi_{23}-\phi_{14})/4} & 0 \\
0 & e^{i\phi_{14}/4}
\end{pmatrix},
\end{equation}
the generic entry $\rho_{ij}$ ($i \neq j$) becomes $|\rho_{ij}|$.
This means that $U_{A}\otimes U_{B}$ turns $\rho_X$ into a density matrix whose entries are real and non-negative.
This property is useful for the calculation of the quantities of interest in this paper - quantum correlations - since they are all invariant under local unitary transformations.

Thus, it is possible to parametrize the generic $X$-state just with five parameters in the following way
\begin{eqnarray}\label{ro new}
\rho_{X} = \frac{1}{4} \left(
\begin{array}{cccc}
1+r+s+c_3
& 0 & 0 & c_1 -c_2 \\
0 & 1+r-s-c_3 & c_1+c_2 & 0 \\
0 & c_1 +c_2 & 1-r+s-c_3
& 0 \\
c_1 -c_2 & 0 & 0 & 1-r-s+c_3
\end{array}
\right).
\label{Xstate}
\end{eqnarray}
Such a writing can be represented through the Bloch normal form \cite{li,yur,cel,gir,nam,sun,kim}
\begin{eqnarray}
\rho=\frac{1}{4}[I\otimes I+\textbf{r}\cdot{\sigma}\otimes I+I\otimes\textbf{s}\cdot{\sigma}+ c_1\sigma^{x}\otimes\sigma^{x} + c_2\sigma^{y}\otimes\sigma^{y} +c_3\sigma^{z}\otimes\sigma^{z} ],
\label{twoqubit}
\end{eqnarray}
with the two Bloch vectors $\textbf{r}=(0,0,r)$ and $\textbf{s}=(0,0,s)$ ($\sigma=(\sigma^{x},\sigma^{y},\sigma^{z}$) are the standard Pauli matrices).
We underline that if \textbf{r}=\textbf{s}=0, $\rho$ becomes the Bell diagonal state.

As a measure of entanglement we shall use Wootter's concurrence (entanglement of formation \cite{wootters} is a monotonically increasing
function of the concurrence), which can be calculated by using the eigenvalues of $\rho\widetilde{\rho}$, where $\widetilde{\rho}= \sigma^y\otimes \sigma^y\rho^*\sigma^y\otimes \sigma^y$.
The eigenvalues of $\rho\widetilde{\rho}$ for the state (\ref{Xstate}) are
\begin{eqnarray}
\lambda_1=\frac{1}{16}(c_1-c_2-\sqrt{(1+c_3)^2-(r+s)^2})^2, \qquad \nonumber
\lambda_2=\frac{1}{16}(c_1-c_2+\sqrt{(1+c_3)^2-(r+s)^2})^2,\nonumber
\end{eqnarray}
\begin{eqnarray}
\lambda_3=\frac{1}{16}(c_1+c_2-\sqrt{(1-c_3)^2-(r-s)^2})^2, \qquad \nonumber
\lambda_4=\frac{1}{16}(c_1+c_2+\sqrt{(1-c_3)^2-(r-s)^2})^2
\end{eqnarray}
and the concurrence is given by
\begin{eqnarray}
C(\rho)=\max\{2\max\{\sqrt{\lambda_1},\sqrt{\lambda_2},\sqrt{\lambda_3},\sqrt{\lambda_4}\}
-\sqrt{\lambda_1}-\sqrt{\lambda_2}-\sqrt{\lambda_3}-\sqrt{\lambda_4},0 \}.
\label{twoqubitconc}
\end{eqnarray}
For fixed $r$ and $s$, the previous states and their corresponding concurrence depend on three parameters.

For two-qubit X states with density matrices of the form (\ref{Xstate}), the quantum discord
can be computed analytically according to the procedure elaborated in Refs. \cite{ali,li}, and shortly described in the following.
The quantum mutual information can be expressed in the form
\begin{equation}
\mathcal{I}(\rho)=S(\rho^A)+S(\rho^B)+u_+\log_2u_+
+u_-\log_2u_-+v_+\log_2v_++v_-\log_2v_-,
\label{mutualinformation}
\end{equation}
with
\begin{equation}S(\rho^A)=1+f(r), \quad S(\rho^B)=1+f(s), \qquad
f(t)=-\frac{1-t}{2}\log_2(1-t)-\frac{1+t}{2}\log_2(1+t),~~~~ 0\leq t\leq 1
\end{equation}
and
\begin{eqnarray}{}
u_\pm=\frac{1}{4}[1-c_3\pm\sqrt{(r-s)^2+(c_1+c_2)^2} ], \quad
v_\pm=\frac{1}{4}[1+c_3\pm\sqrt{(r+s)^2+(c_1-c_2)^2} ].
\end{eqnarray}
being the two eigenvalues of $\rho_X$ in Eq. (\ref{Xstate}).

After performing the von Neumann measurement $B_i$, $i=0,1$
for the subsystem $B$,  one obtains the ensemble $\{\rho_i, p_i\}$
and then the classical correlations $\mathcal{C}(\rho)$ can be evaluated by
\begin{equation}
\mathcal{C}(\rho) = \sup_{B_i} \, \mathcal{I} (\rho|B_i)
   = S(\rho^A)- \min_{B_i} S (\rho|B_i),        \label{Eq:CC}
\end{equation}
where
\begin{equation}
S(\rho|B_i)=p_0S(\rho_0)+p_1S(\rho_1).
\label{conditionalentropy}
\end{equation}
According to Refs. \cite{ali,li} the minimum of the quantum conditional entropy (\ref{conditionalentropy}) has to be taken over the following expressions:
\begin{eqnarray}
S_1   = -\frac{1+r+s+c_3}{4}\log_2\frac{1+r+s+c_3}{2(1+s)}
   -\frac{1-r+s-c_3}{4}\log_2\frac{1-r+s-c_3}{2(1+s)}  \nonumber\\
     -\frac{1+r-s-c_3}{4}\log_2\frac{1+r-s-c_3}{2(1-s)}
      -\frac{1-r-s+c_3}{4}\log_2\frac{1-r-s+c_3}{2(1-s)},      \label{Eq:S1}
\end{eqnarray}
\begin{equation}
S_2=1+f(\sqrt{r^2+c_1^2}),      \label{Eq:S2}
\end{equation}
\begin{equation}
S_3=1+f(\sqrt{r^2+c_2^2}).      \label{Eq:S3}
\end{equation}
Finally, Li \cite{li} formulated the following

{ \bf Theorem}:
For any state $\rho$ of the form (\ref{Xstate}), the classical correlations of $\rho$ are given by
\begin{eqnarray}
\label{proposition1}
\mathcal{C}(\rho)= S(\rho^A) - \min\{S_1, S_2, S_3\},
\end{eqnarray}
where $S_1, S_2, S_3$ are defined by Eqs. (\ref{Eq:S1}), (\ref{Eq:S2}), (\ref{Eq:S3}) respectively.
The quantum discord is then given by
\begin{eqnarray}\label{qd}
D(\rho)=\mathcal{I}(\rho)-\mathcal{C}(\rho),
\end{eqnarray}
with $\mathcal{I}(\rho)$ given by Eq. (\ref{mutualinformation}).

The same results are obtained by generalizing von Neumann measurements to POVM \cite{ali}.

\section{Hamiltonian model}
\label{sec:3}

We are interested in studying analytically the time evolution of the quantum discord exhibited by a two-interacting-spin-1/2 system when the Hamiltonian governing the dynamics is time-dependent.
Such a task turns out to be hard for two important reasons.
First, as highlighted in the previous section, analytical expressions for the QD are possible only for X-states and, in general, an initial X-state of two spin 1/2's does not keep the X-structure in successive time-instants.
To this end, thus, we look for a non-trivial and physically relevant two-spin-1/2 time-dependent model accomplishing the request to leave the X-structure of an initial state unchanged.
That is, a model for which an initial X-state evolves non-trivially in time remaining an X-state.
Secondly, we know that non-trivial exactly solvable time-dependent models of interacting qudits are very rare and difficult to be found.
However, quite recently an interesting time-dependent model of two interacting spin-1/2's, with a clear and transparent physical meaning, satisfying the two previous requirements has been introduced \cite{grimess}.
Such a model, examined in detail and solved in \cite{grimess}, reads:
\begin{equation} \label{Hamiltonian}
{H} =
\hbar\omega_{1}(t)\hat{\sigma}_{1}^{z}+\hbar\omega_{2}(t)\hat{\sigma}_{2}^{z}+\gamma_{xx}\hat{\sigma}_{1}^{x}\hat{\sigma}_{2}^{x}
+\gamma_{yy}\hat{\sigma}_{1}^{y}\hat{\sigma}_{2}^{y}+\gamma_{zz}\hat{\sigma}_{1}^{z}\hat{\sigma}_{2}^{z}+\gamma_{xy}\hat{\sigma}_{1}^{x}\hat{\sigma}_{2}^{y}
+\gamma_{yx}\hat{\sigma}_{1}^{y}\hat{\sigma}_{2}^{x} \
\end{equation}
where $\hat{\sigma}_{i}^{x}$, $\hat{\sigma}_{i}^{y}$ and $\hat{\sigma}_{i}^{z}$ ($i=1,2$) are the Pauli matrices and all the parameters may be thought as time-dependent.
The matrices are represented in the standard two-spin basis ordered as follows $\{ \ket{++},\ket{+-},\ket{-+},\ket{--} \}$ $(\hat{\sigma}^z\ket{\pm}=\pm\ket{\pm})$.

According to the symmetry-based argumentations reported in Ref. \cite{grimess}, the time evolution operator, solution of the Schr\"odinger equation $i\hbar\dot{U}=HU$, may be formally put in the following form
\begin{equation}\label{Total time ev op}
U =
\begin{pmatrix}
|a_{+}|e^{i \Phi_{a}^{+}} & 0 & 0 & |b_{+}|e^{i \Phi_{b}^{+}} \\
0 & |a_{-}|e^{i \Phi_{a}^{-}} & |b_{-}|e^{i \Phi_{b}^{-}} & 0 \\
0 & -|b_{-}|e^{-i \Phi_{b}^{'-}} & |a_{-}|e^{-i \Phi_{a}^{'-}} & 0 \\
-|b_{+}|e^{-i \Phi_{b}^{'+}} & 0 & 0 & |a_{+}|e^{-i \Phi_{a}^{'+}} \\
\end{pmatrix},
\end{equation}
where
\begin{align}
 \Phi_{a/b}^{\pm} = \phi_{a/b}^{\pm} \mp \dfrac{\gamma_{zz}}{\hbar} t, \qquad
 \Phi_{a/b}^{' \pm} = \phi_{a/b}^{\pm} \pm \dfrac{\gamma_{zz}}{\hbar} t.
\end{align}
The condition $U(0)=\mathbb{1}$ is satisfied by putting $a_\pm(0)=1$ and $b_\pm(0)=0$.
It is worth to note that $a_\pm(t)\equiv|a_\pm(t)|e^{i\phi_a^\pm(t)}$ and $b_\pm(t)\equiv|b_\pm(t)|e^{i\phi_b^\pm(t)}$ are the time-dependent parameters of the two evolution operators
\begin{equation}
U_{\pm} = e^{\mp i \gamma_{zz} t / \hbar}
\begin{pmatrix}
|a_{\pm}|e^{i \phi_{a}^{\pm}} & |b_{\pm}|e^{i \phi_{b}^{\pm}} \\
-|b_{\pm}|e^{-i \phi_{b}^{\pm}} & |a_{\pm}|e^{-i \phi_{a}^{\pm}}
\end{pmatrix},
\end{equation}
solutions of two independent dynamical Cauchy problems of fictitious single spin-1/2, namely $i\hbar \dot{U}_\pm = H_\pm U_\pm$, $U_\pm(0)=\mathbb{1}_\pm$, with
\begin{equation}
H_{\pm}=
\begin{pmatrix}
\Omega_{\pm}(t) & \Gamma_{\pm} \\
\Gamma_{\pm}^{*} & -\Omega_{\pm}(t)
\end{pmatrix}
\pm \gamma_{zz} \mathbb{1},
\end{equation}
where
\begin{equation} \label{Omega+- and Gamma+-}
\begin{aligned}
\Omega_{\pm}(t) = \hbar [\omega_{1}(t) \pm \omega_{2}(t)], \qquad
\Gamma_{\pm} = (\gamma_{xx} \mp \gamma_{yy}) - i (\pm \gamma_{xy} + \gamma_{yx}).
\end{aligned}
\end{equation}
Thus, it means that the solution of the dynamical problems of the two interacting spin-1/2's is traced back to the solution of two independent problems, each one of single (fictitious) spin-1/2.

It is possible to convince oneself that $\rho(0)=\rho_X$ evolves keeping the X-structure at any time instant.
Thus, preparing our two-spin system in the X-state \eqref{X-state}, at any subsequent time instant $\rho(t)=U(t)\rho(0)U^\dagger(t)$ is still an X-state, $U(t)$ being the operator defined in Eq. \eqref{Total time ev op}.
The proof is based on the following argument reported in Ref. \cite{grimess}.

The property exhibited by $\rho(t)$ is due to the special structure of the time evolution operator which, in turn, is determined by the symmetry properties of the Hamiltonian.
The $C_2$-symmetry with respect to the $z$-direction, possessed by the Hamiltonian, indeed, causes the existence of two dynamically invariant Hilbert subspaces related to the two eigenvalues of the constant of motion $\hat{\sigma}_1^z\hat{\sigma}_2^z$.
Thus, every state which does not mix the two subspaces at the initial time instant, like the X-states, will keep this property at any following time instant.
This fact implies that, being able to calculate analytically the quantum discord for a generic X-state, we are able to calculate exactly its general time-dependent expression for our model.
Such a general analytical expression will depend on the four parameters $a_\pm(t)$ and $b_\pm(t)$.
In case of exactly scenarios, it means that we have the analytical form of these two parameters, so that we would get an explicit time-dependent expression of the quantum discord.

The explicit expressions of $a_\pm(t)$ and $b_\pm(t)$ depend on the specific time-dependences of the Hamiltonian parameters.
Although, as shown before, the dynamical problem of the two spins may be converted into two independent problems of single spin-1/2, we know that we are  not able to find the analytical solution of the time-dependent Schr\"odinger equation for a spin-1/2 subjected to a generic time-dependent Hamiltonian (that is for generic time-dependences of the Hamiltonian parameters).
Therefore, the knowledge of specific exactly solvable time-dependent scenarios for a single spin-1/2 becomes crucial.
In Ref. \cite{grimess}, the authors present two new classes of time-dependent exactly solvable single-spin-1/2 models and they show how such models can be exploited to construct exactly solvable time-dependent scenarios of the two-interacting-spin system.
Precisely, they considered only the two local magnetic fields acting upon the two spins time-dependent and they showed that, by the knowledge of two exactly solvable time-dependent two-level-system models it is possible to deduce four time-dependent scenarios of the two-interacting qudit model, reported in Appendix \ref{App A}, for which we may construct explicitly the time-evolution operator in Eq. \eqref{Total time ev op}.

Thus, for such exactly solvable time-dependent models of the two-spin system we are able to calculate the explicit form of a generic state and, in particular for an X-state, the explicit time evolution of the quantum discord.
In the next section we analyse the time-dependent quantum discord related to specific initial X-states considered in Appendix \ref{App B}.

\section{Time evolution of quantum discord}
\label{sec:4}

Next we describe the temporal evolution of the quantum discord, using the formalism developed in the previous Section, by taking different initial states for both models considered in this paper (see Appendix \ref{App A} and \ref{App B}).

\begin{figure}
\resizebox{1\textwidth}{!}{%
  \includegraphics{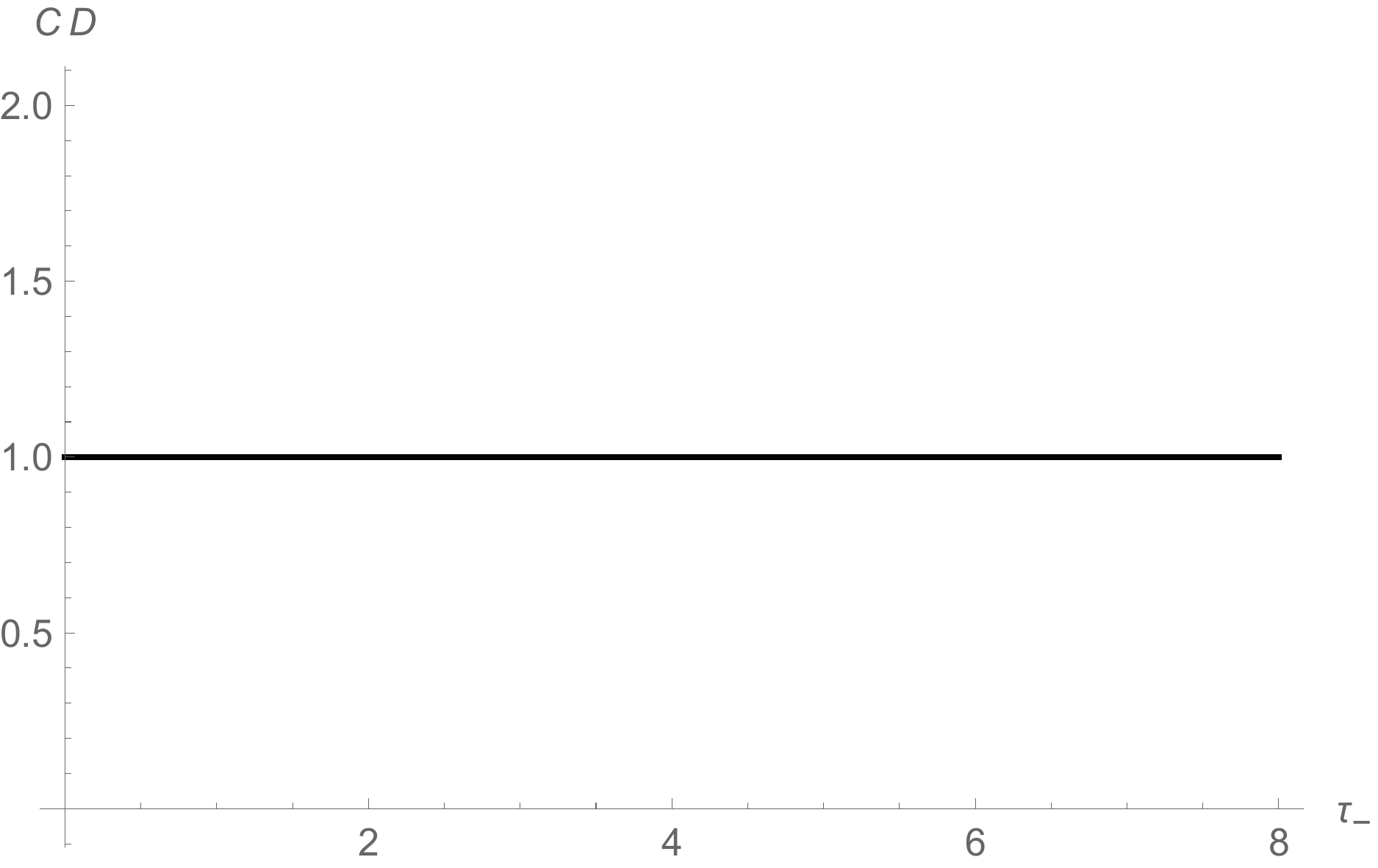}
   \includegraphics{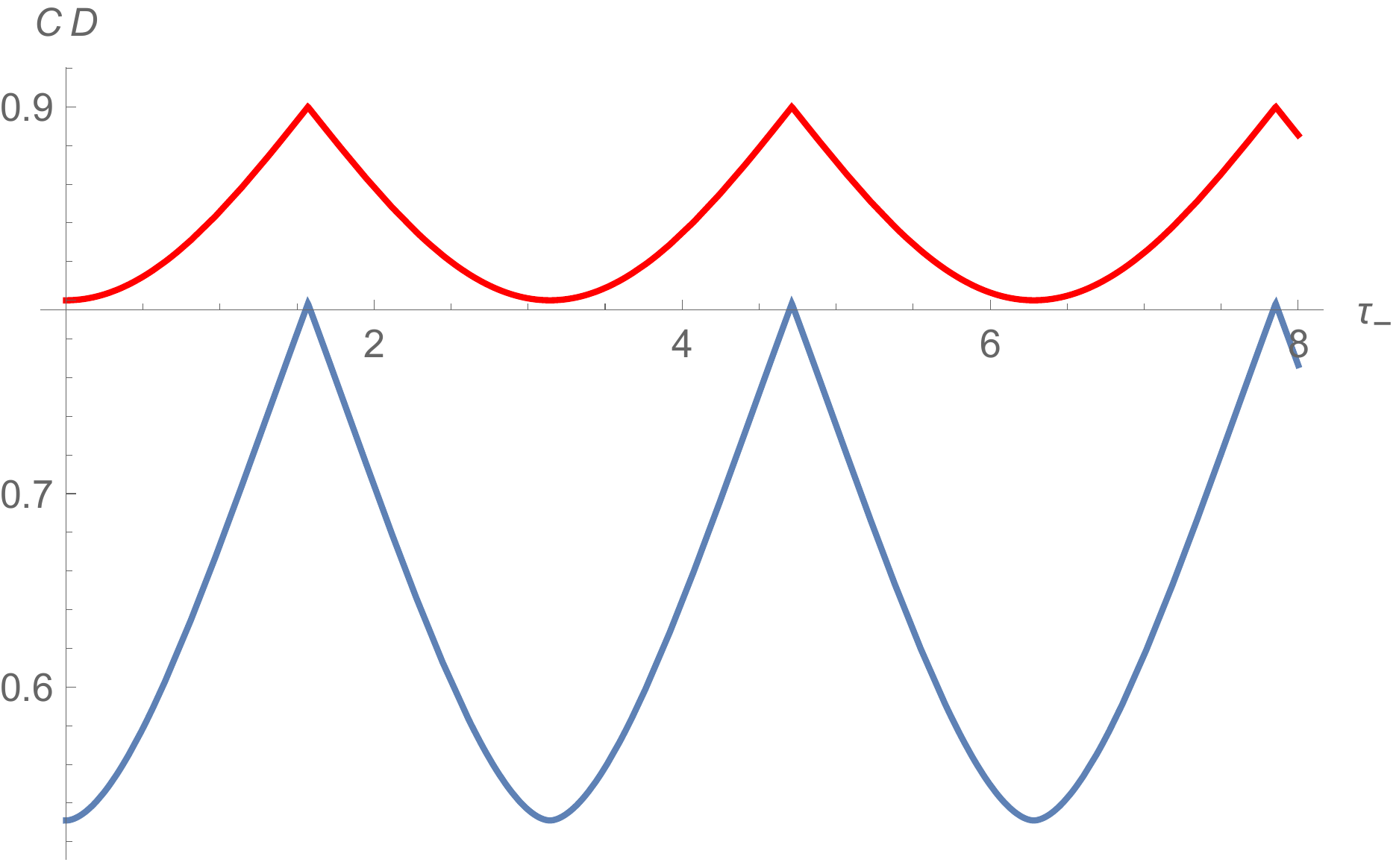}
    \includegraphics{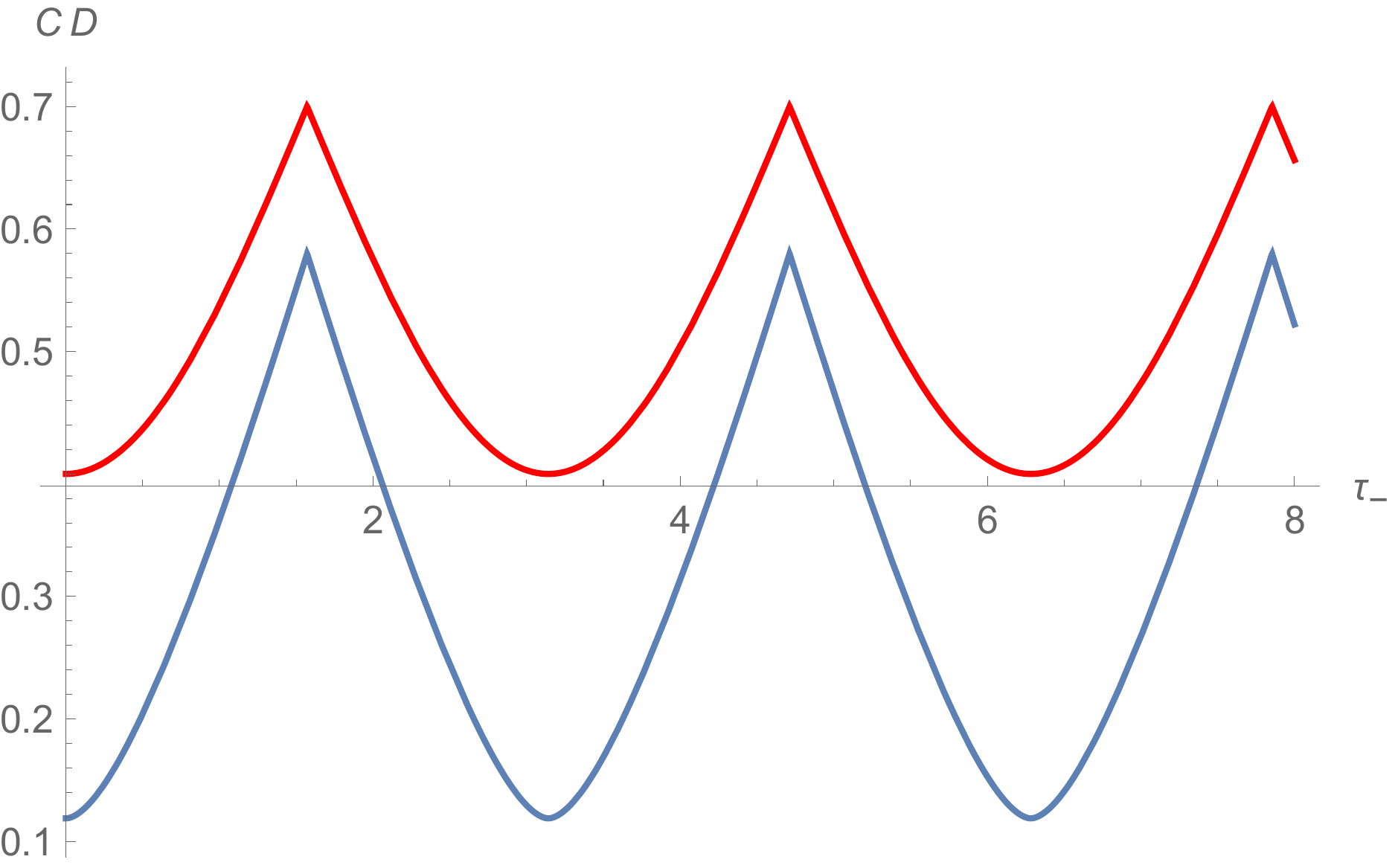}
    }
\resizebox{1\textwidth}{!}  {%
   \includegraphics{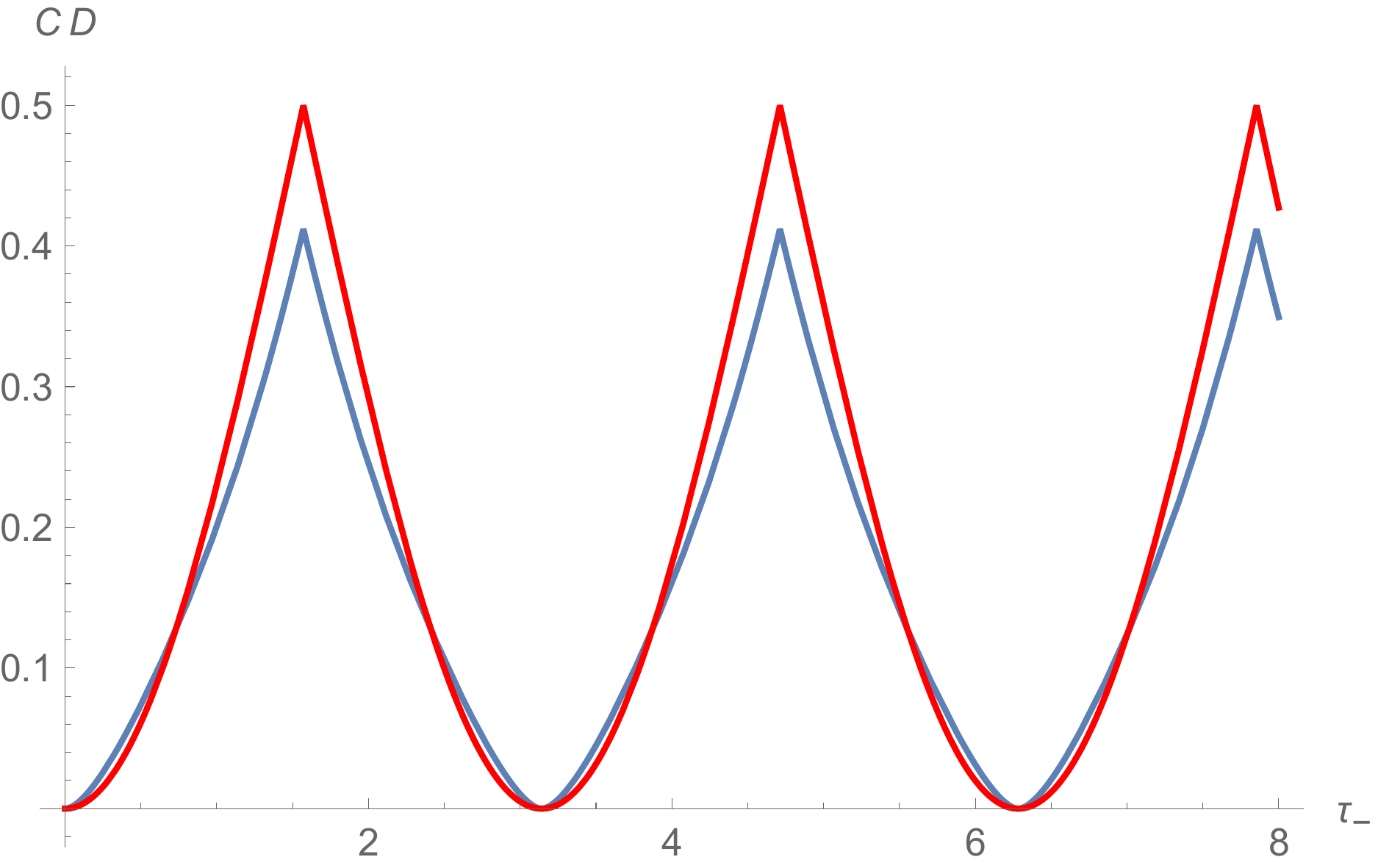}
    \includegraphics{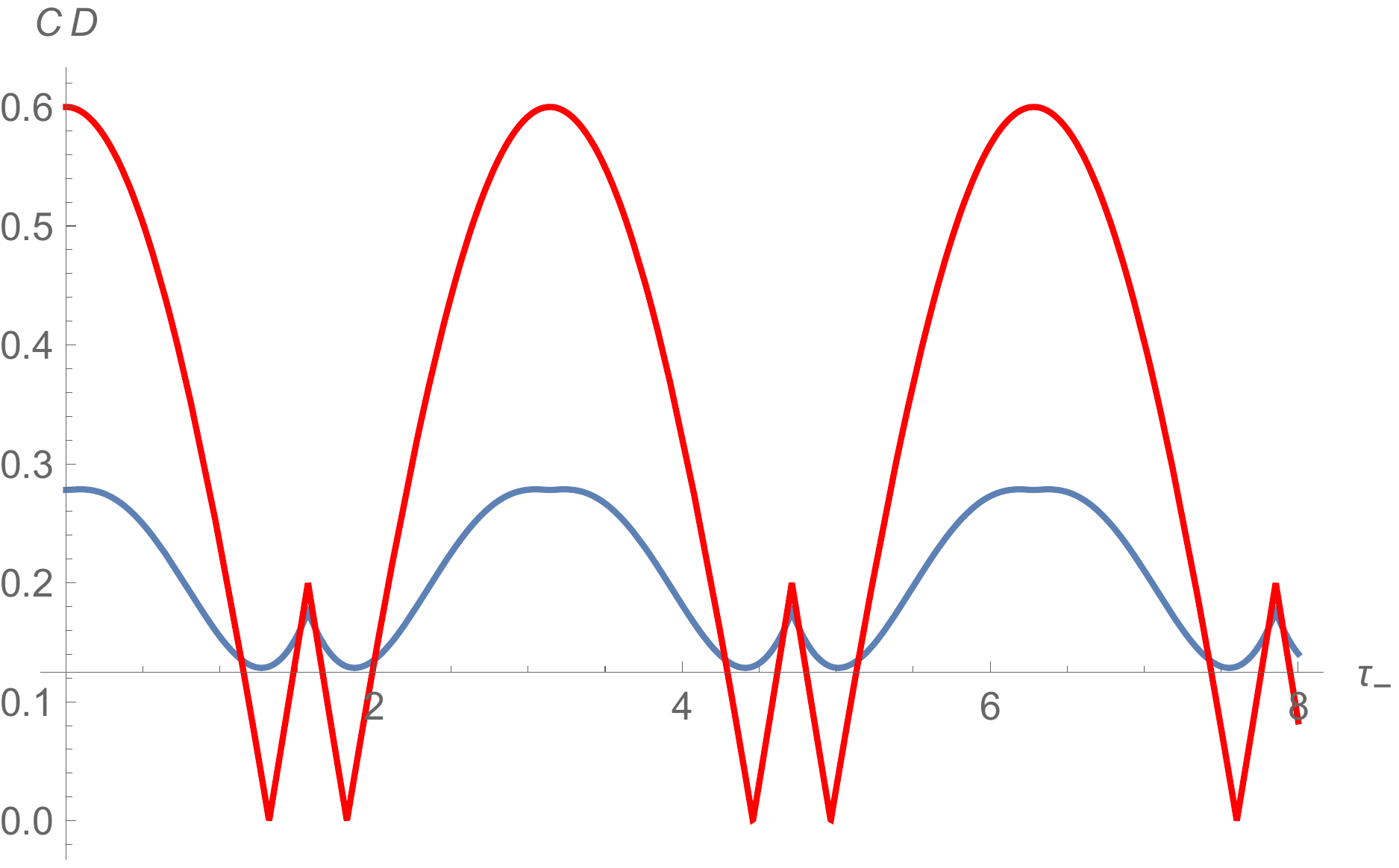}
     \includegraphics{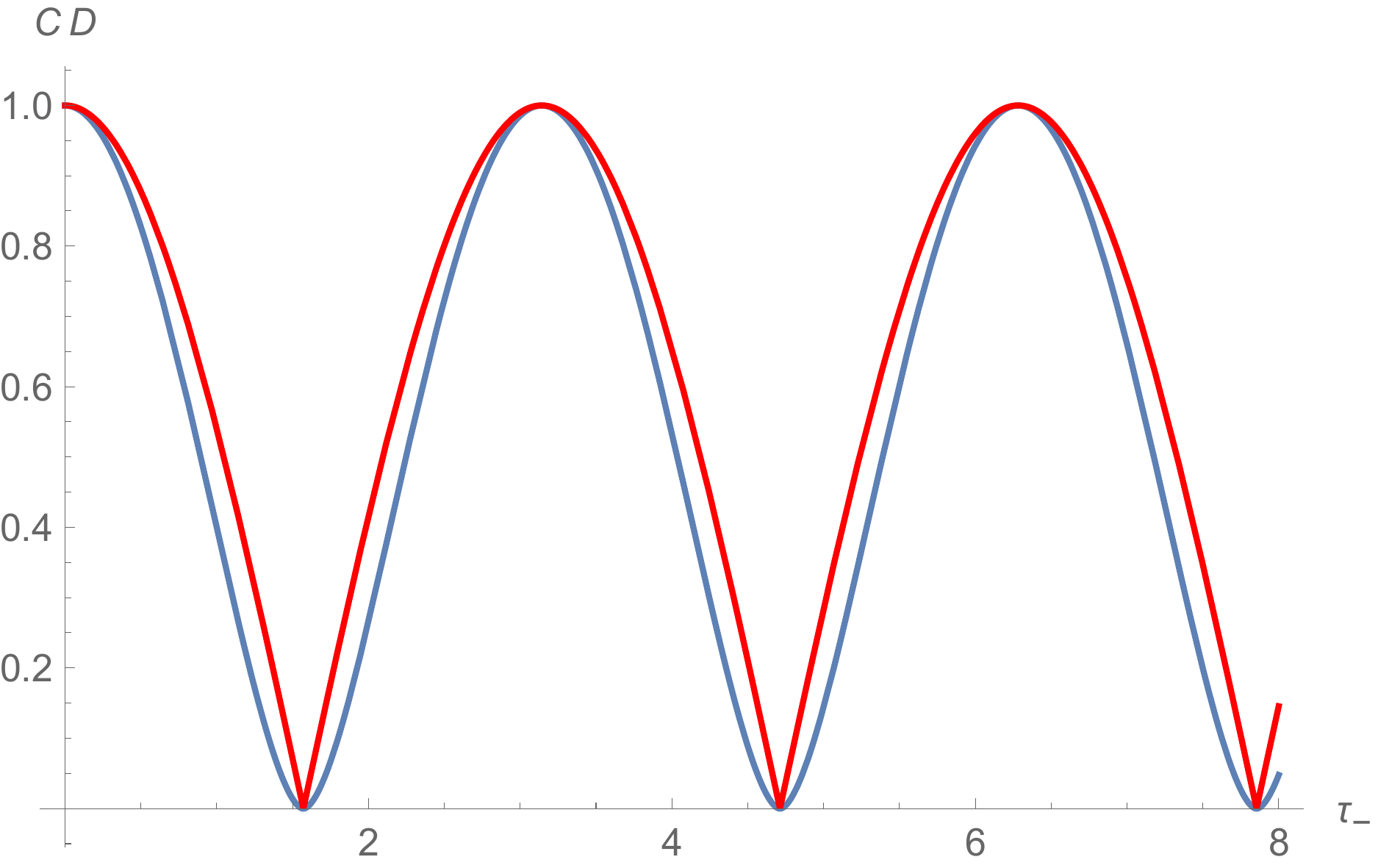}
}
\caption{Quantum concurrence $C$ (red) and quantum discord $D$ (blue) versus time $t$, for various values of the mixing parameter: $p=0,0.1,0.3,0.5,0.8,1,$ respectively, for $\omega_{1}=\omega_{2}=0$, for initial state (\ref{mixt1})-up.}
\label{fig:1}       
\end{figure}

\begin{figure}
\resizebox{1\textwidth}{!}{%
  \includegraphics{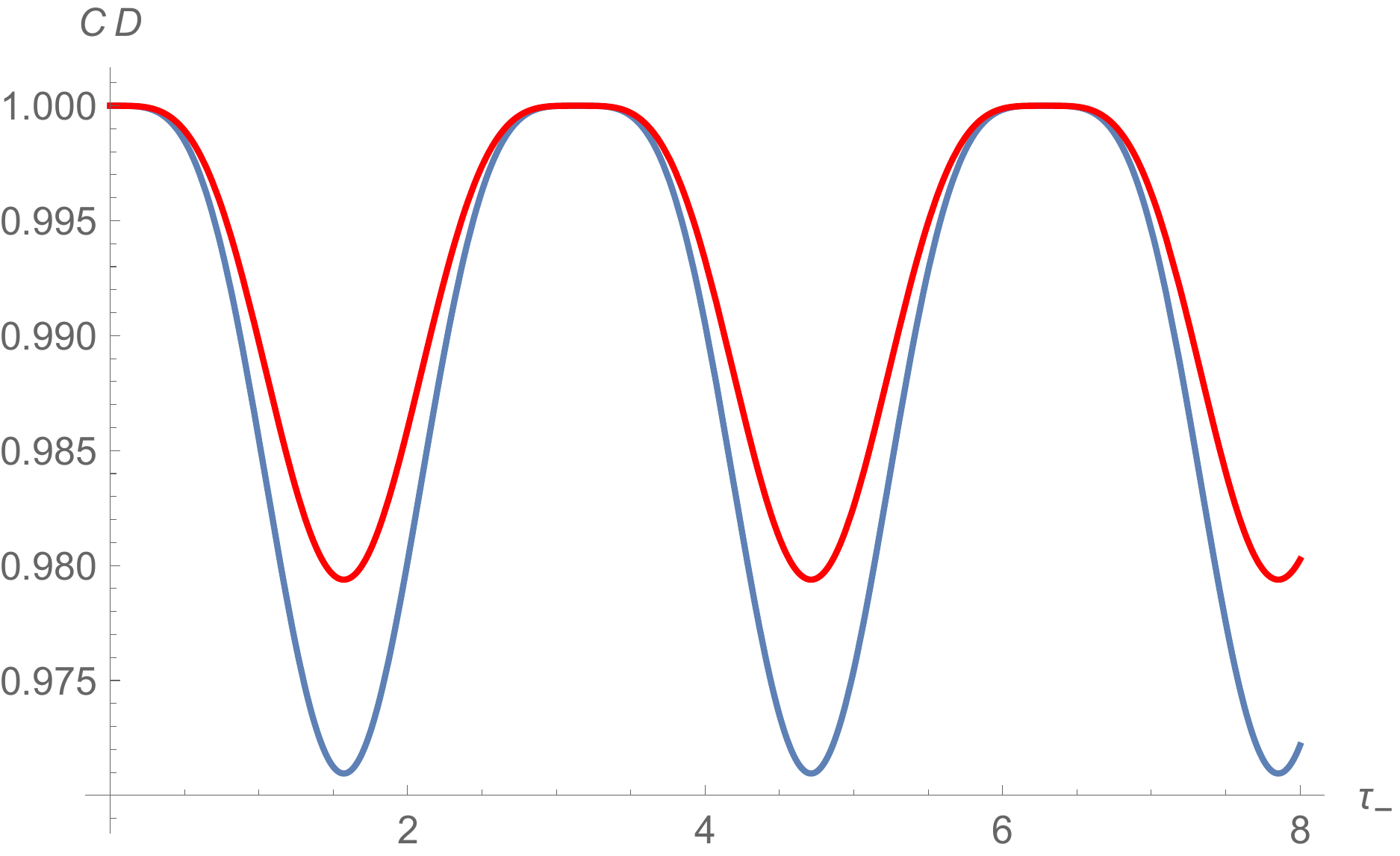}
   \includegraphics{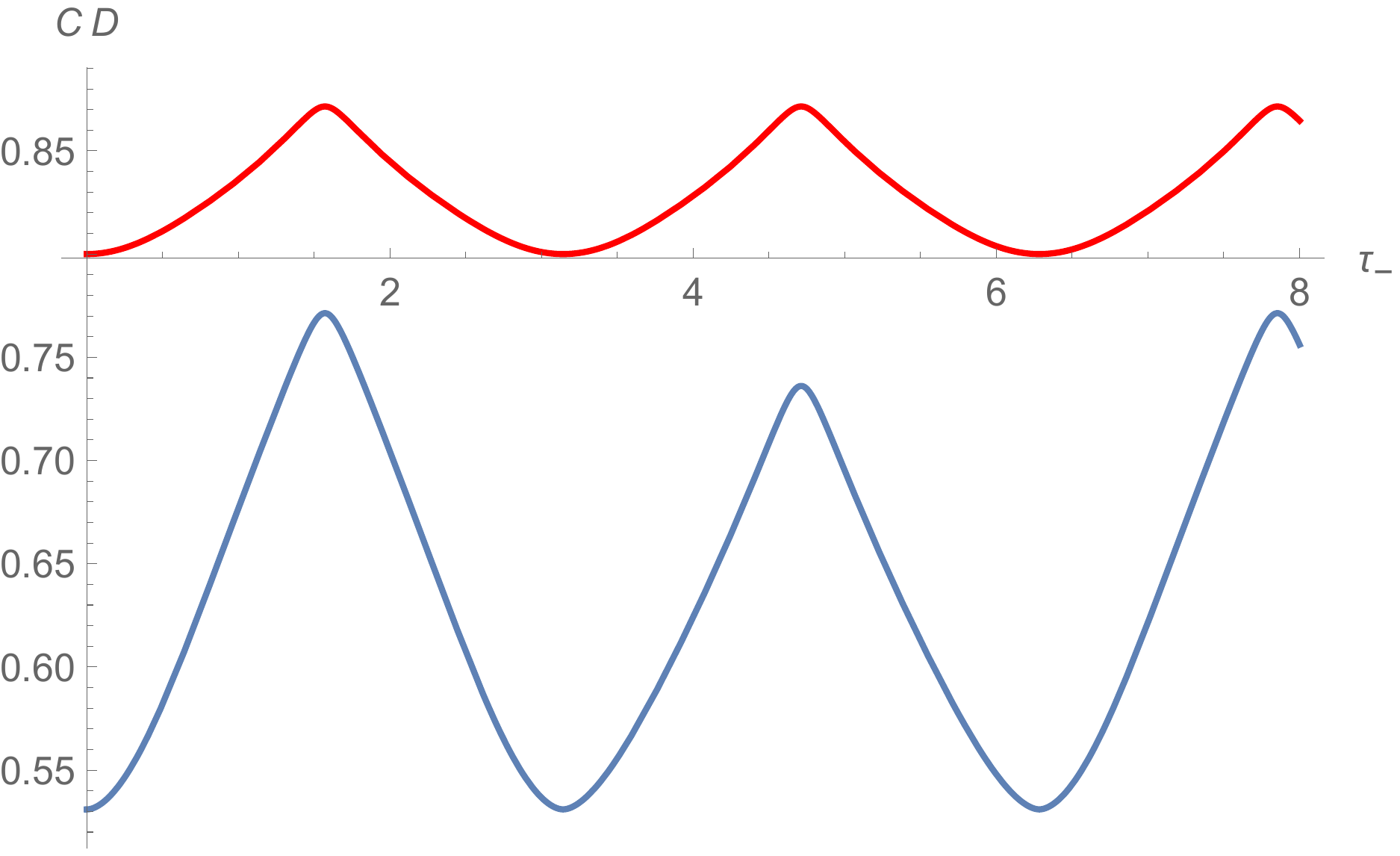}
    \includegraphics{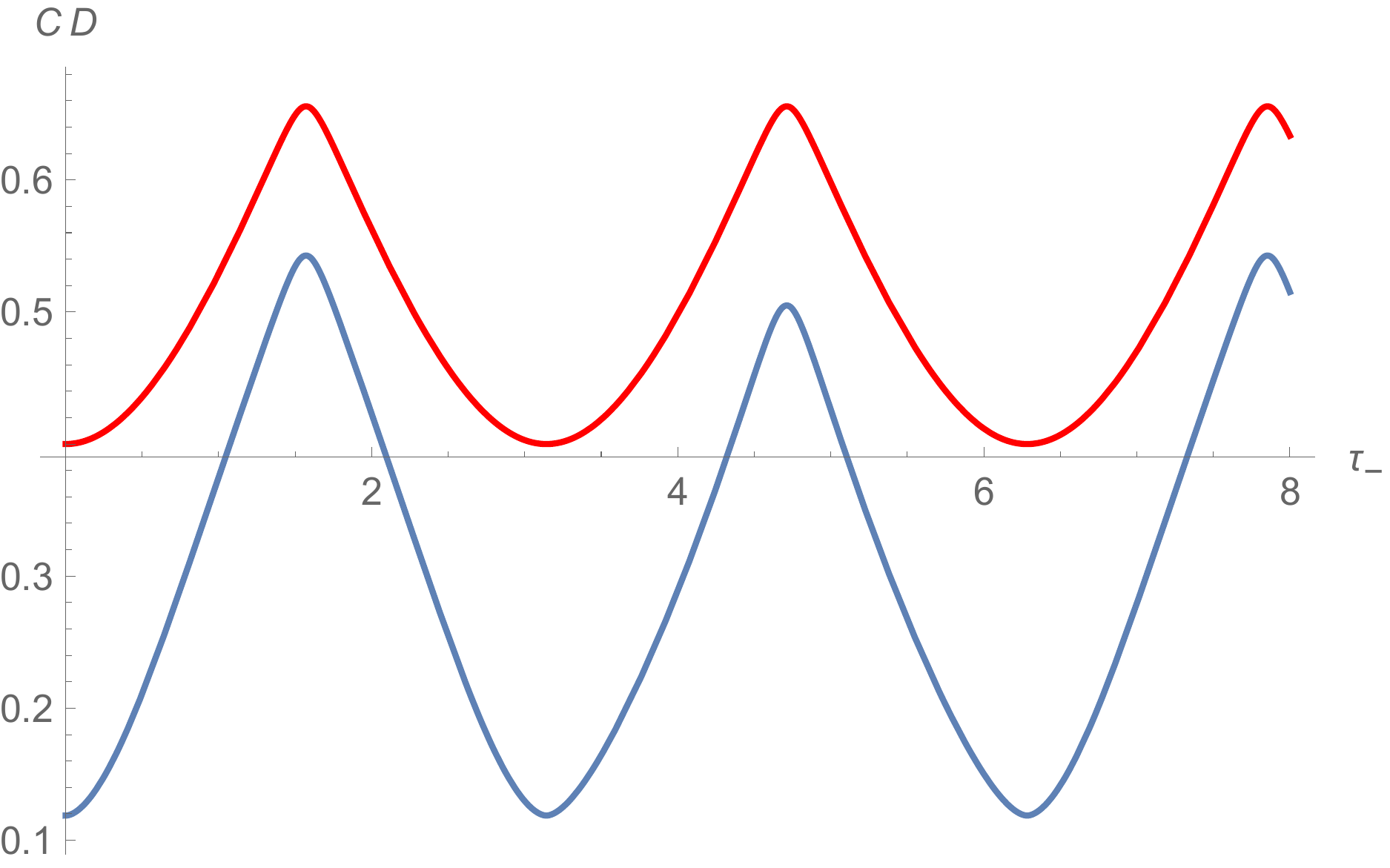}
    }
\resizebox{1\textwidth}{!}  {%
   \includegraphics{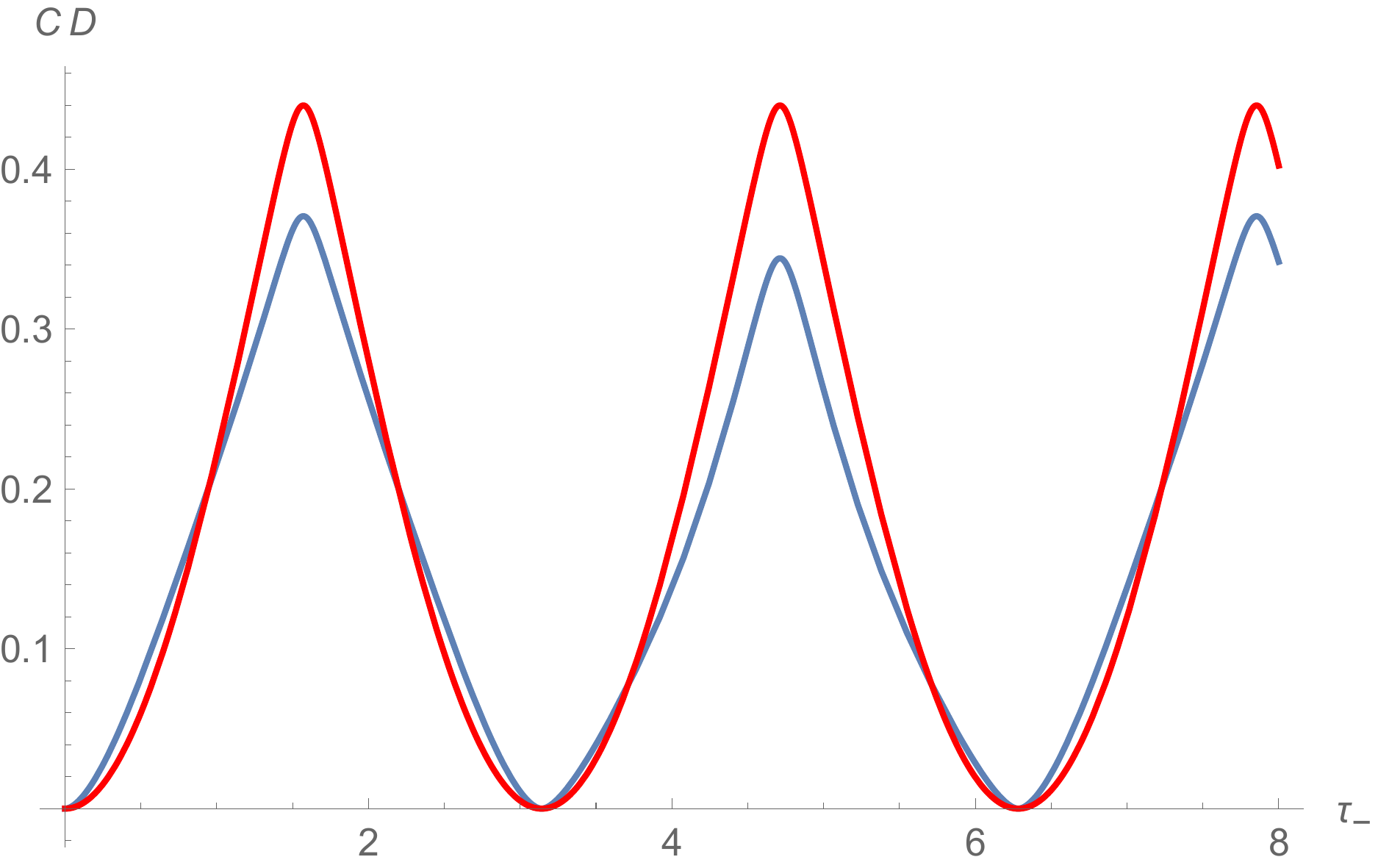}
    \includegraphics{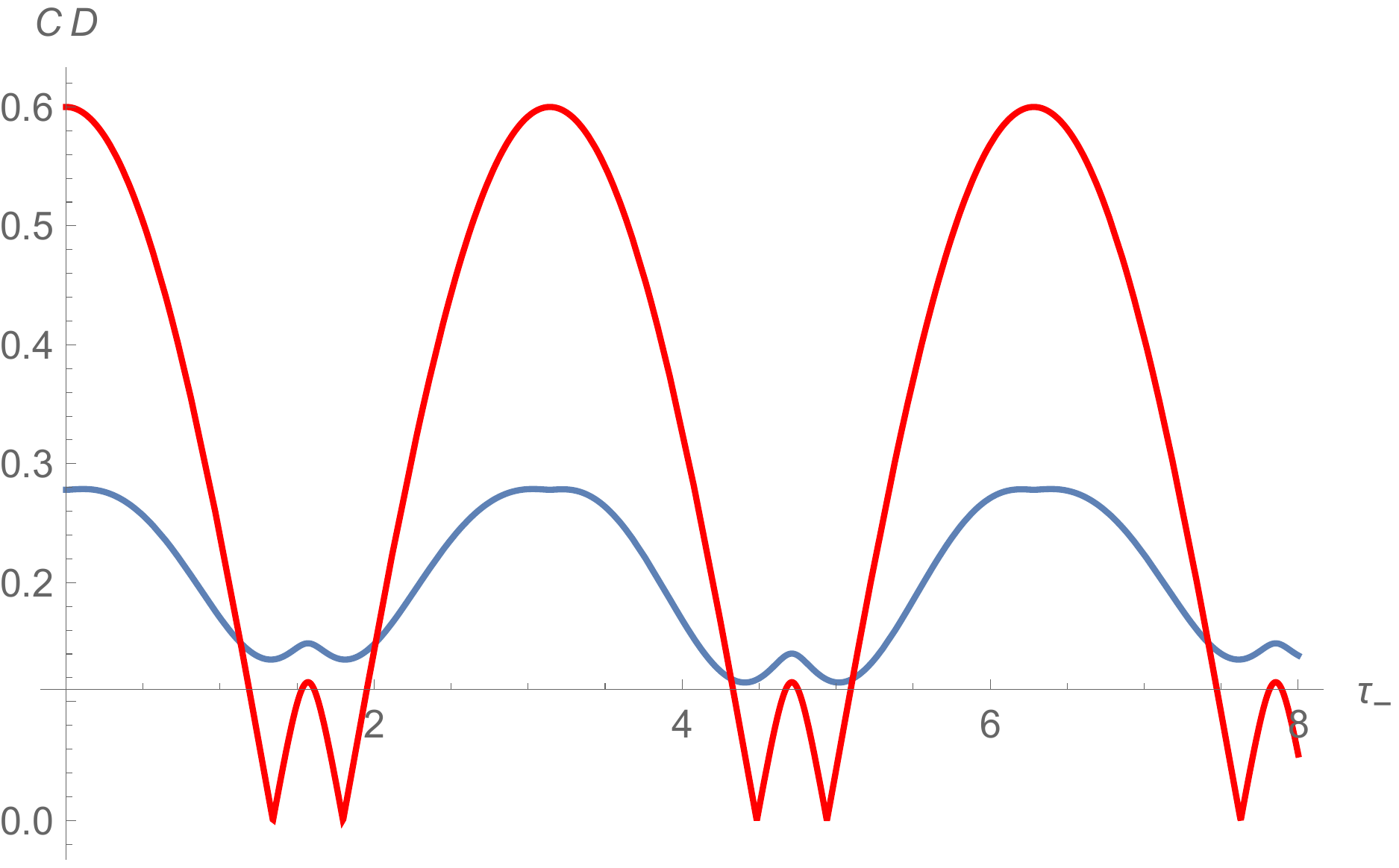}
     \includegraphics{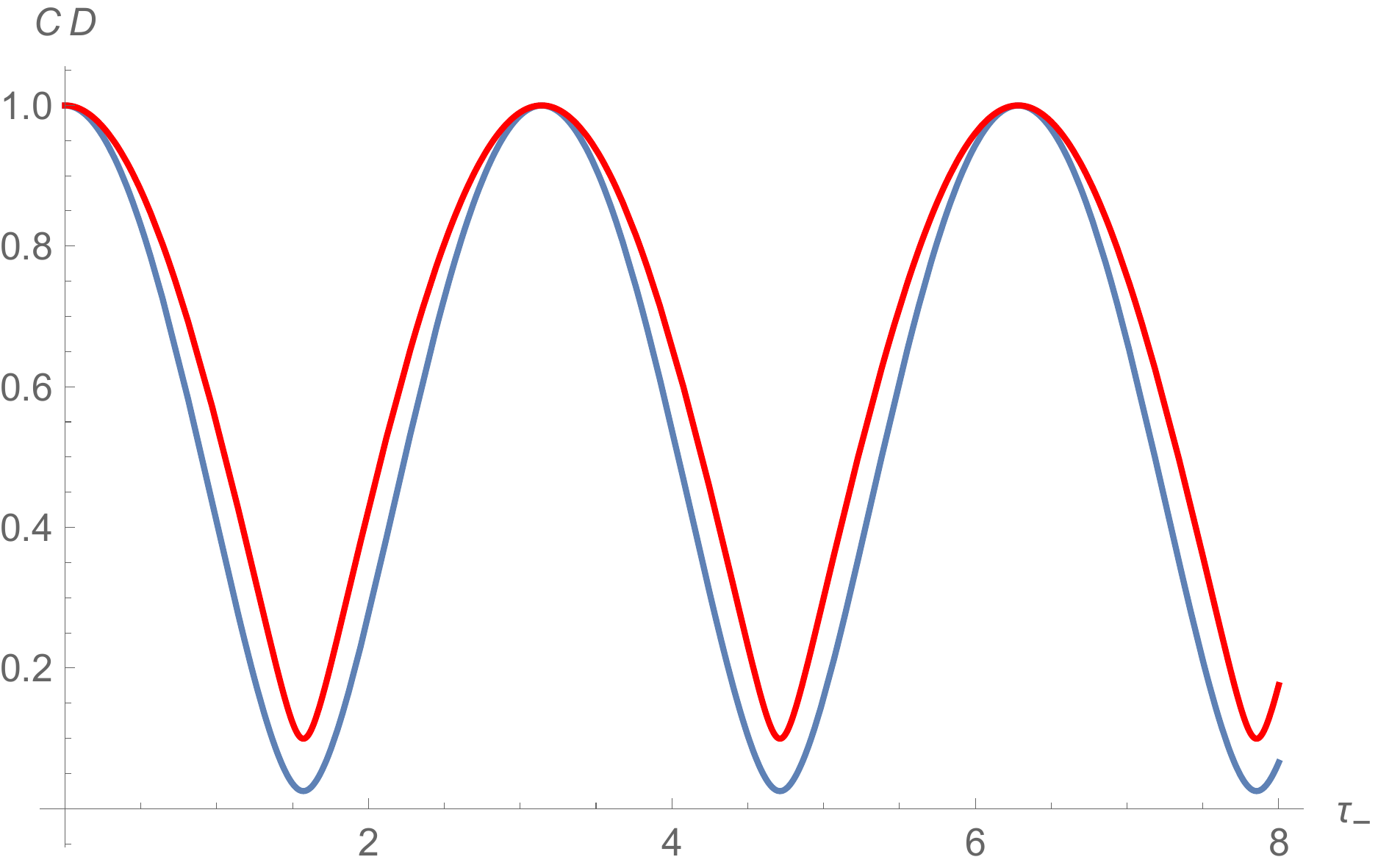}
}
\caption{Quantum concurrence $C$ (red) and quantum discord $D$ (blue) versus time $t$, for various values of the mixing parameter: $p=0,0.1,0.3,0.5,0.8,1,$ respectively, for constant magnetic fields, with $\Omega_{+}=0.1c$, for initial state (\ref{mixt1})-up.}
\label{fig:2}       
\end{figure}

\begin{figure}
\resizebox{1\textwidth}{!}{%
  \includegraphics{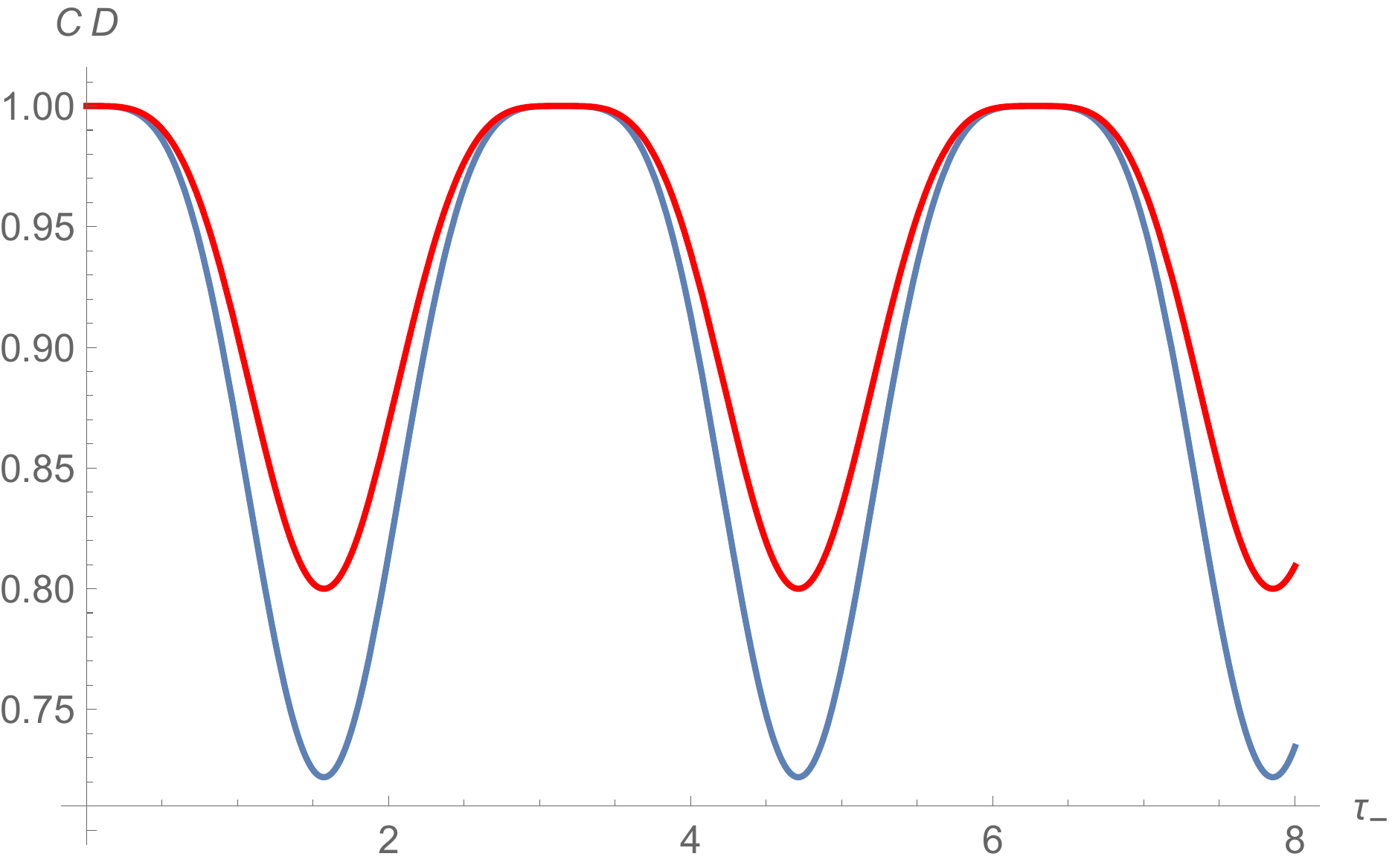}
   \includegraphics{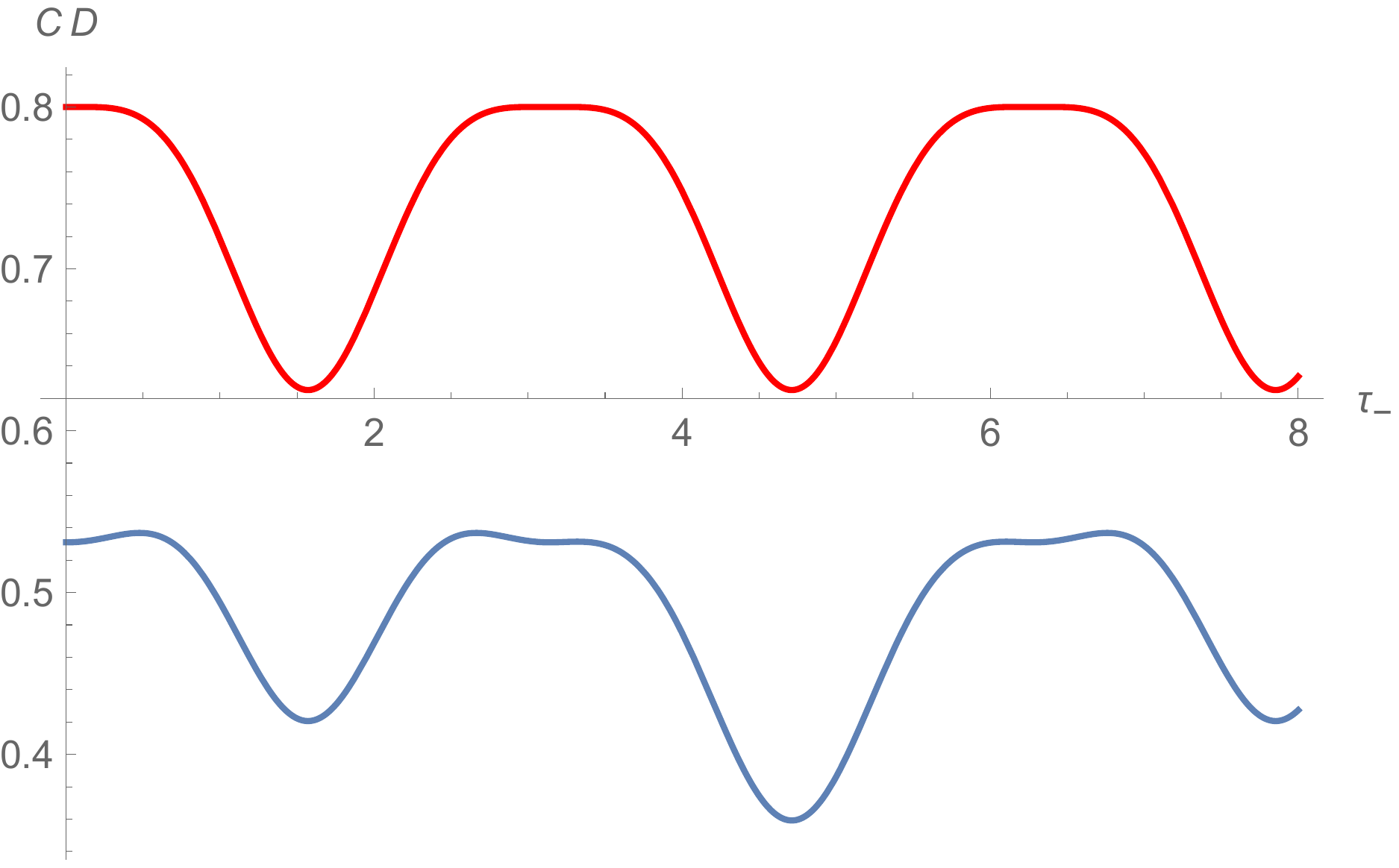}
    \includegraphics{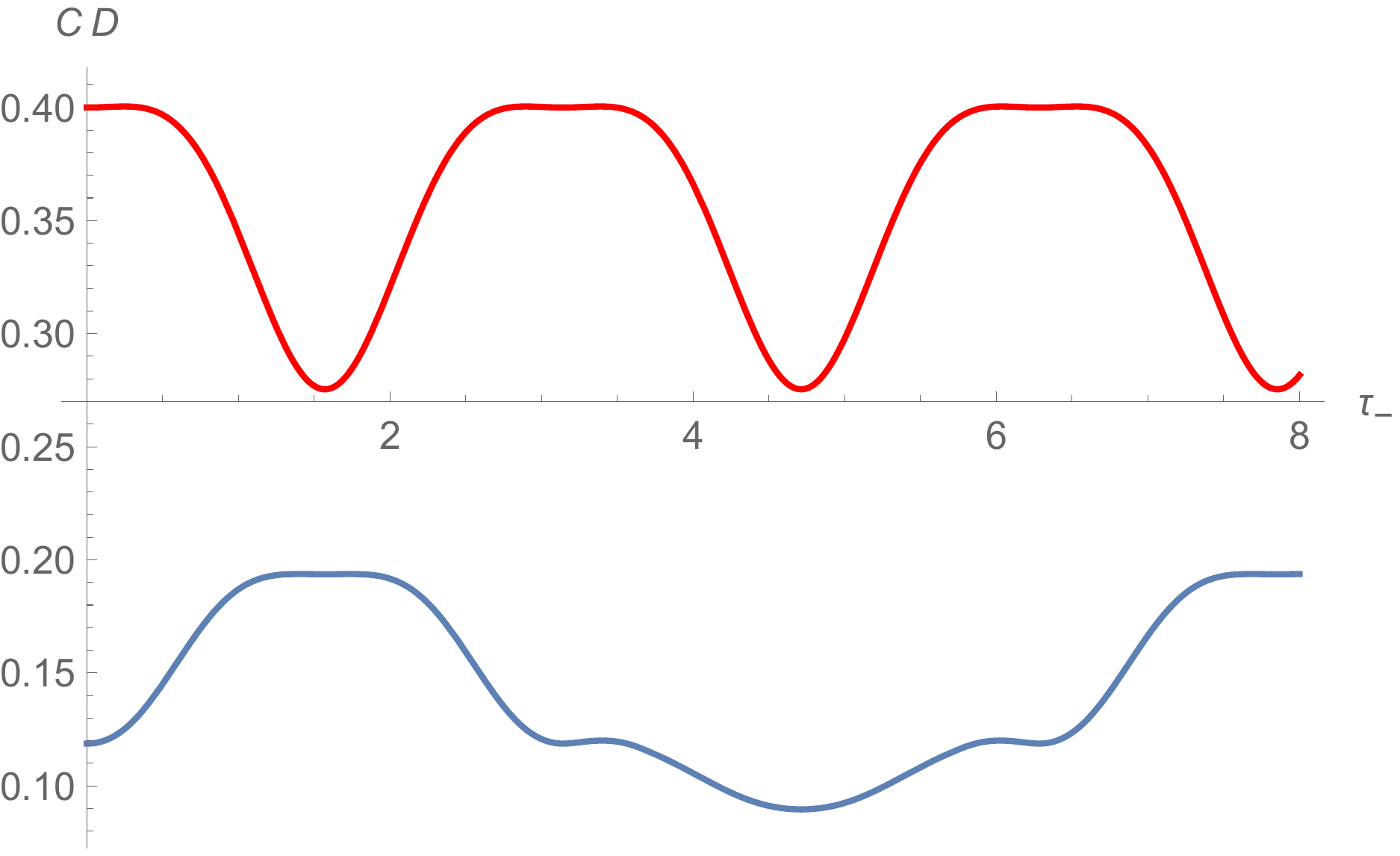}
    }
\resizebox{1\textwidth}{!}  {%
   \includegraphics{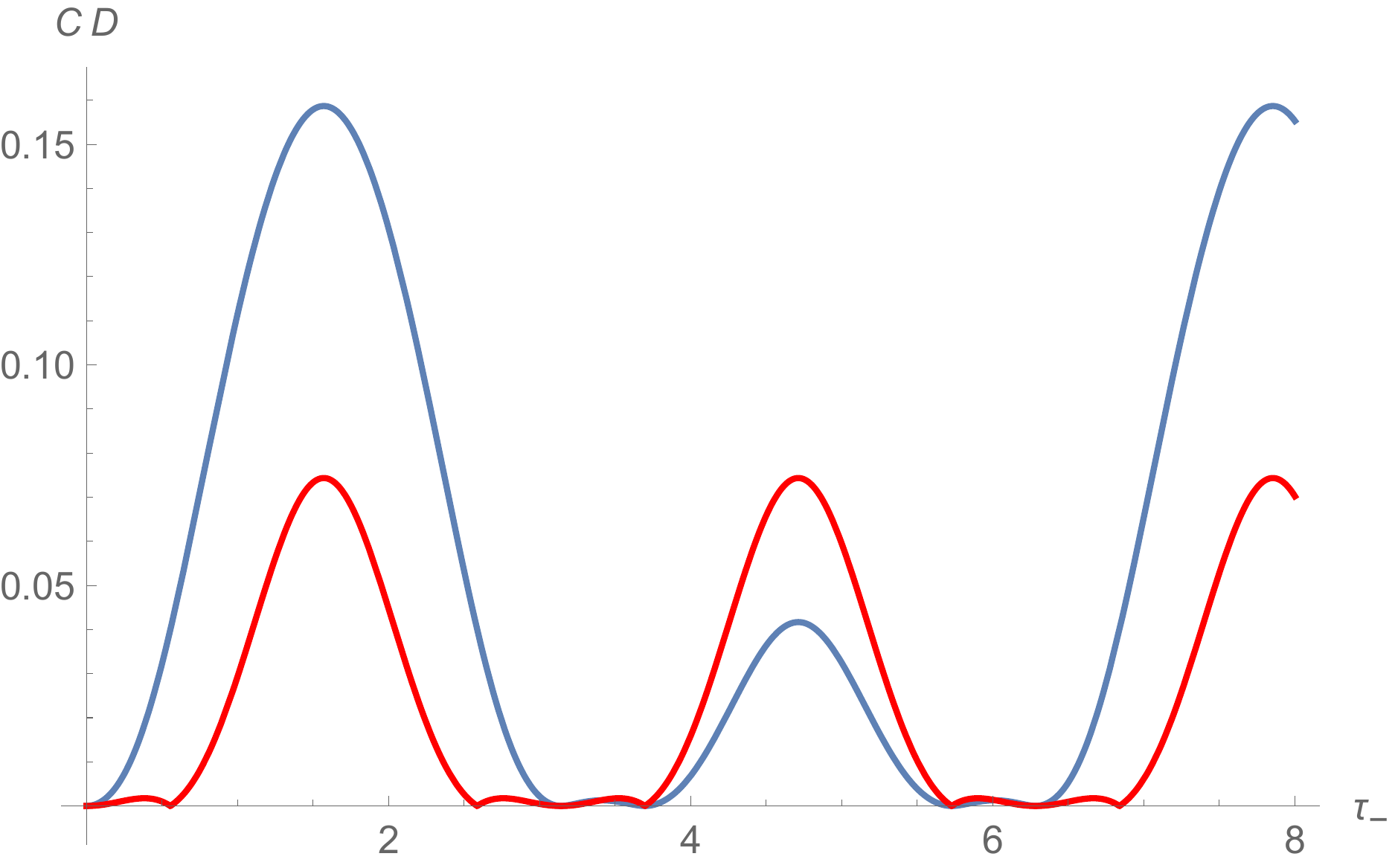}
    \includegraphics{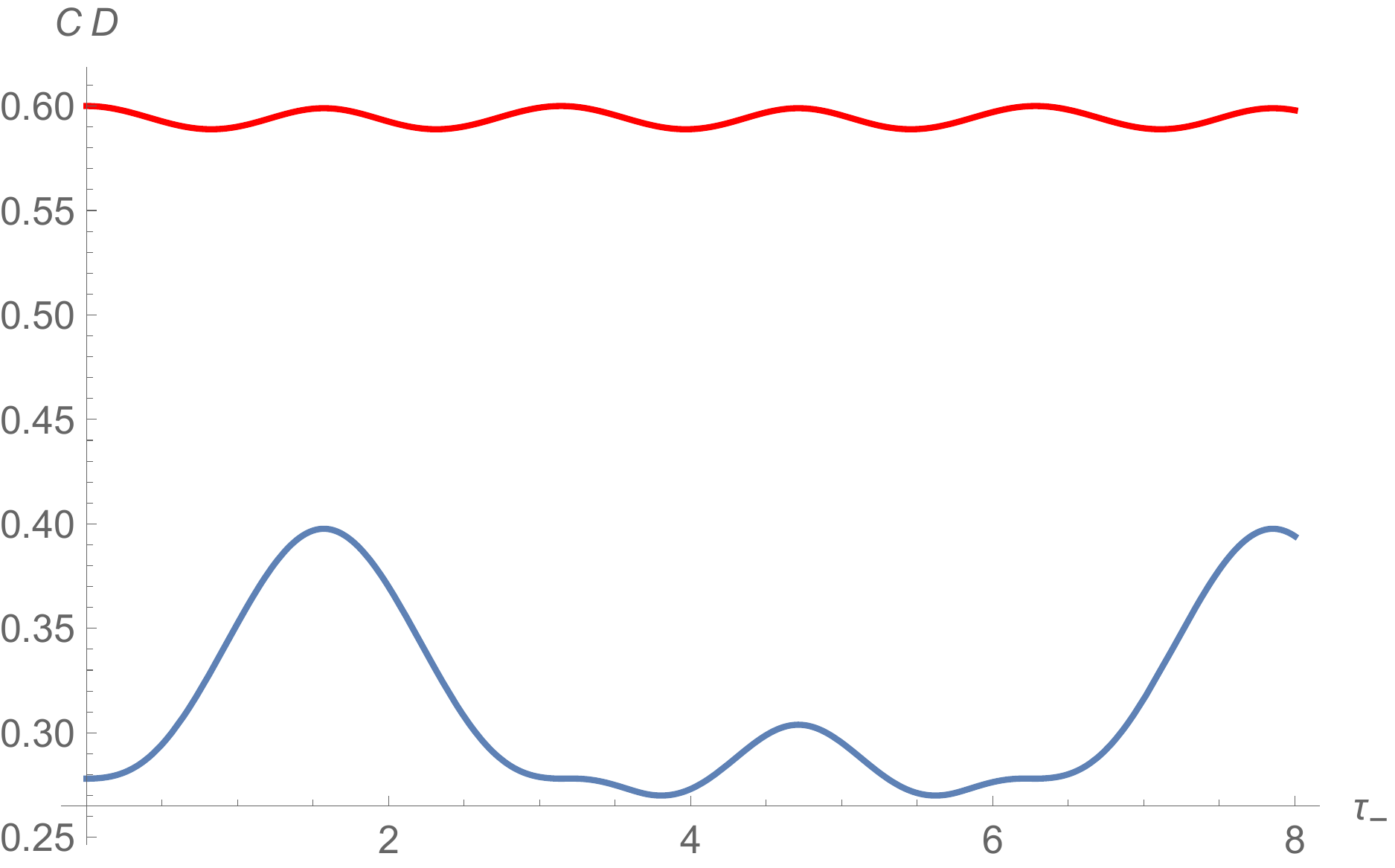}
     \includegraphics{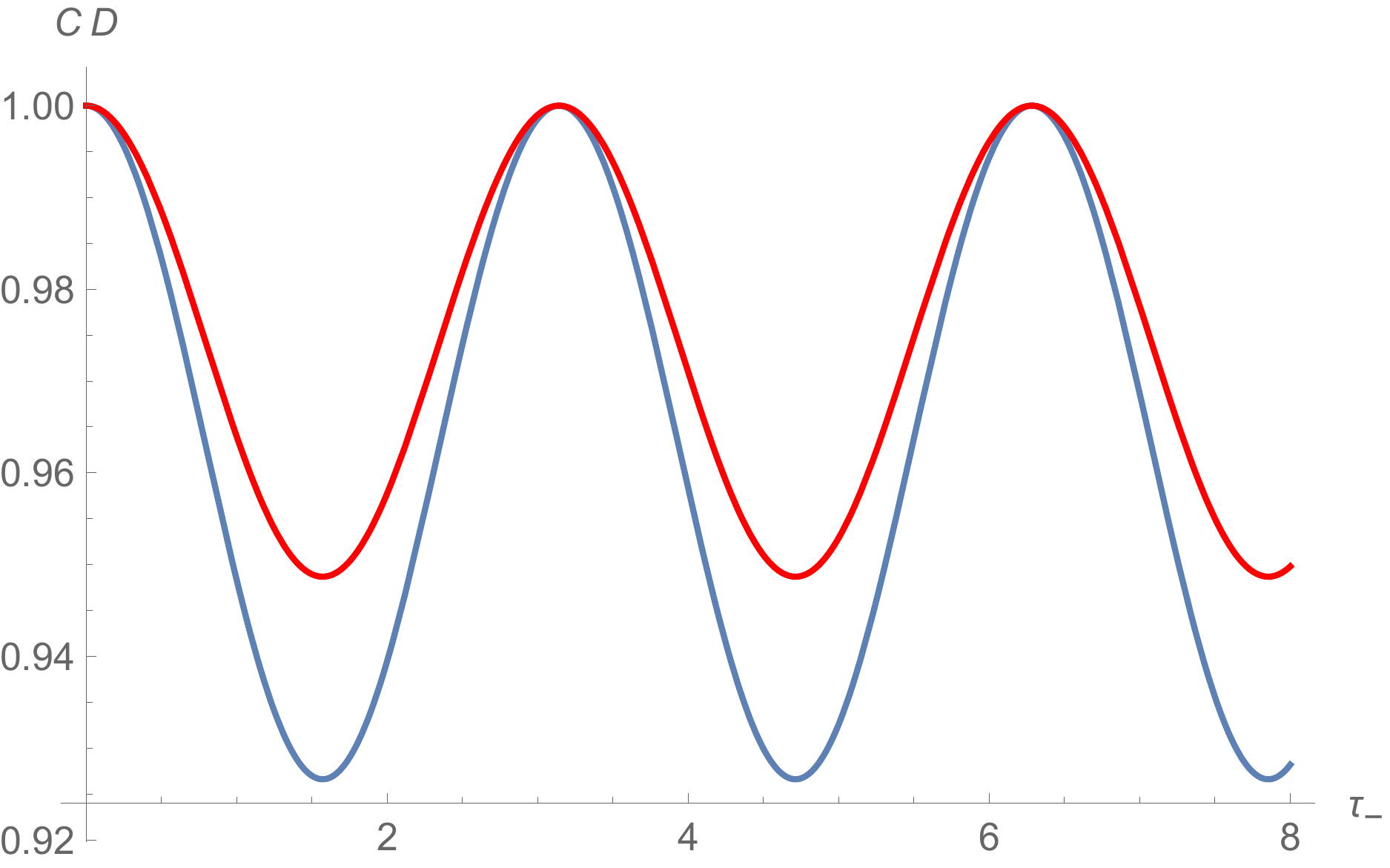}
}
\caption{Quantum concurrence $C$ (red) and quantum discord $D$ (blue) versus time $t$, for various values of the mixing parameter: $p=0,0.1,0.3,0.5,0.8,1,$ respectively, for constant magnetic fields, with $\Omega_{+}=3c$, for initial state (\ref{mixt1})-up.}
\label{fig:3}       
\end{figure}

\begin{figure}
\resizebox{1\textwidth}{!}{%
  \includegraphics{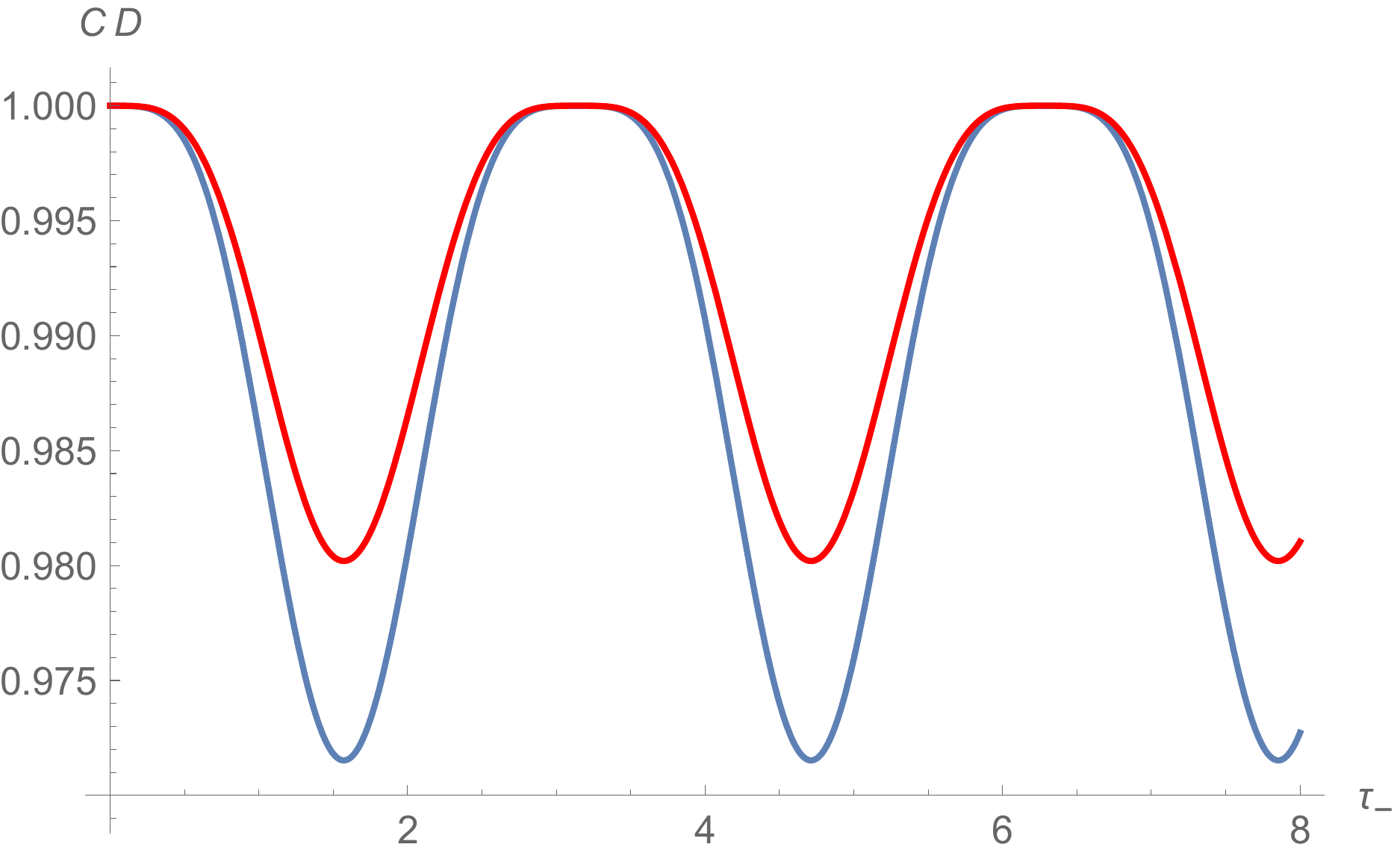}
   \includegraphics{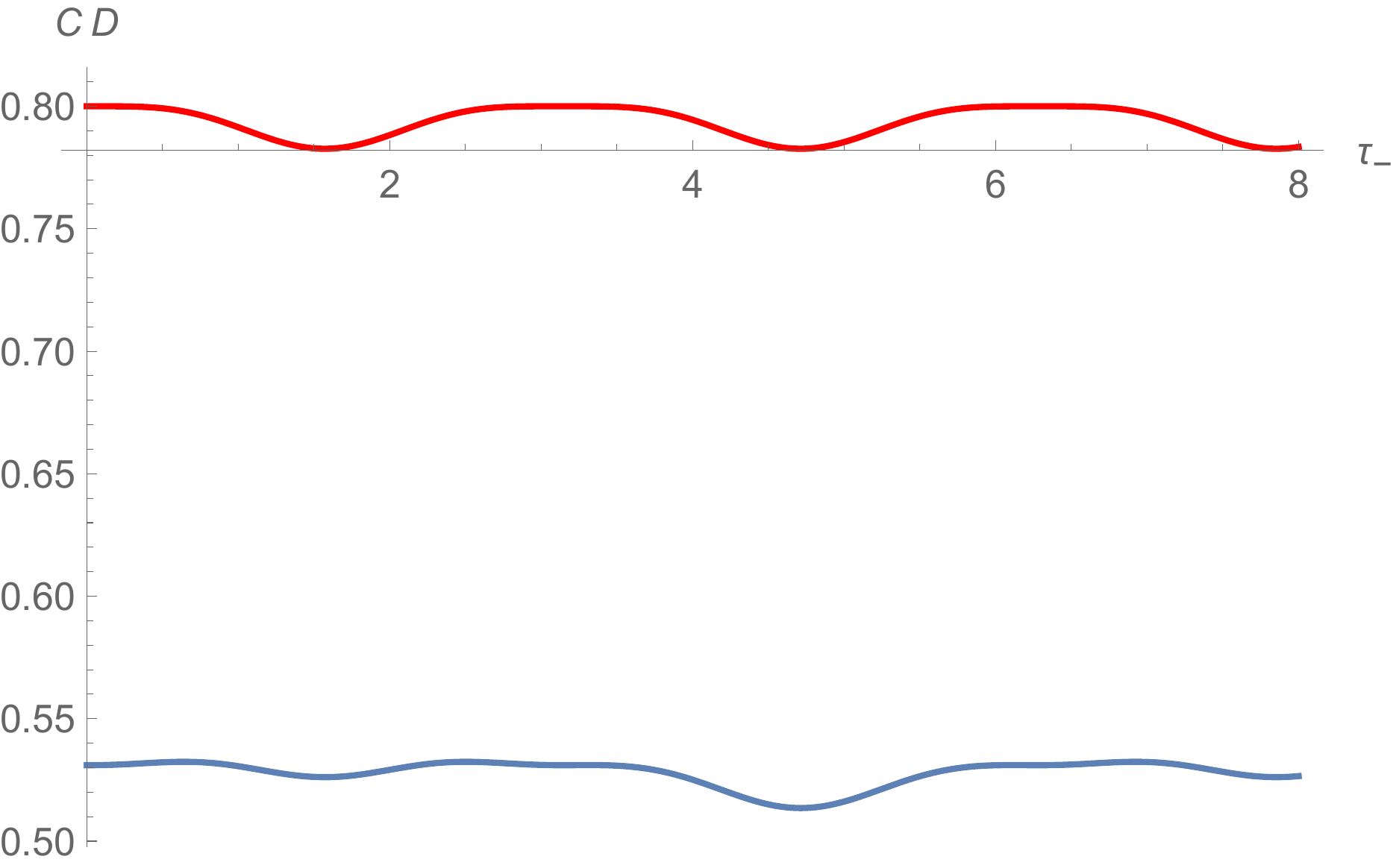}
    \includegraphics{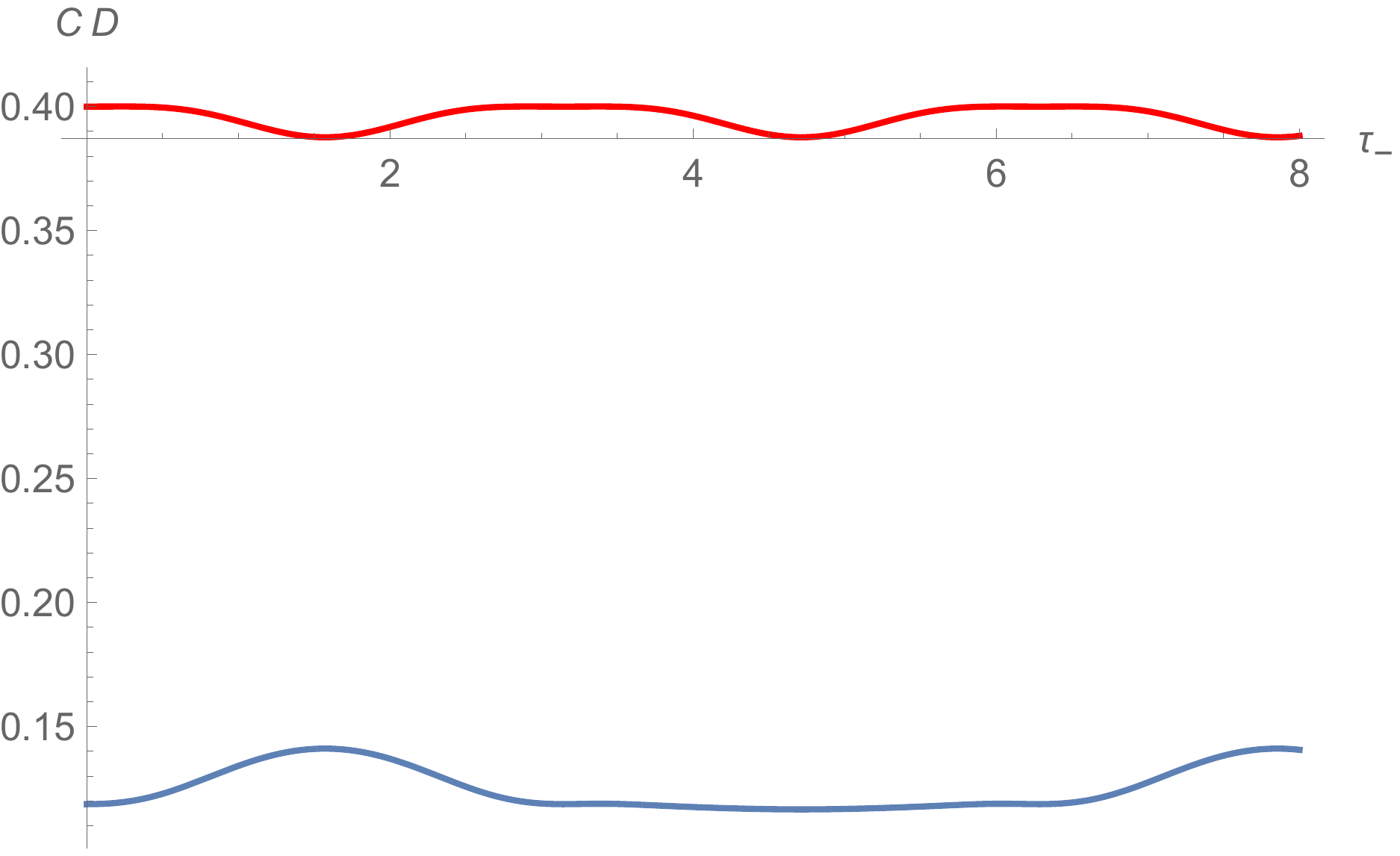}
    }
\resizebox{1\textwidth}{!}  {%
   \includegraphics{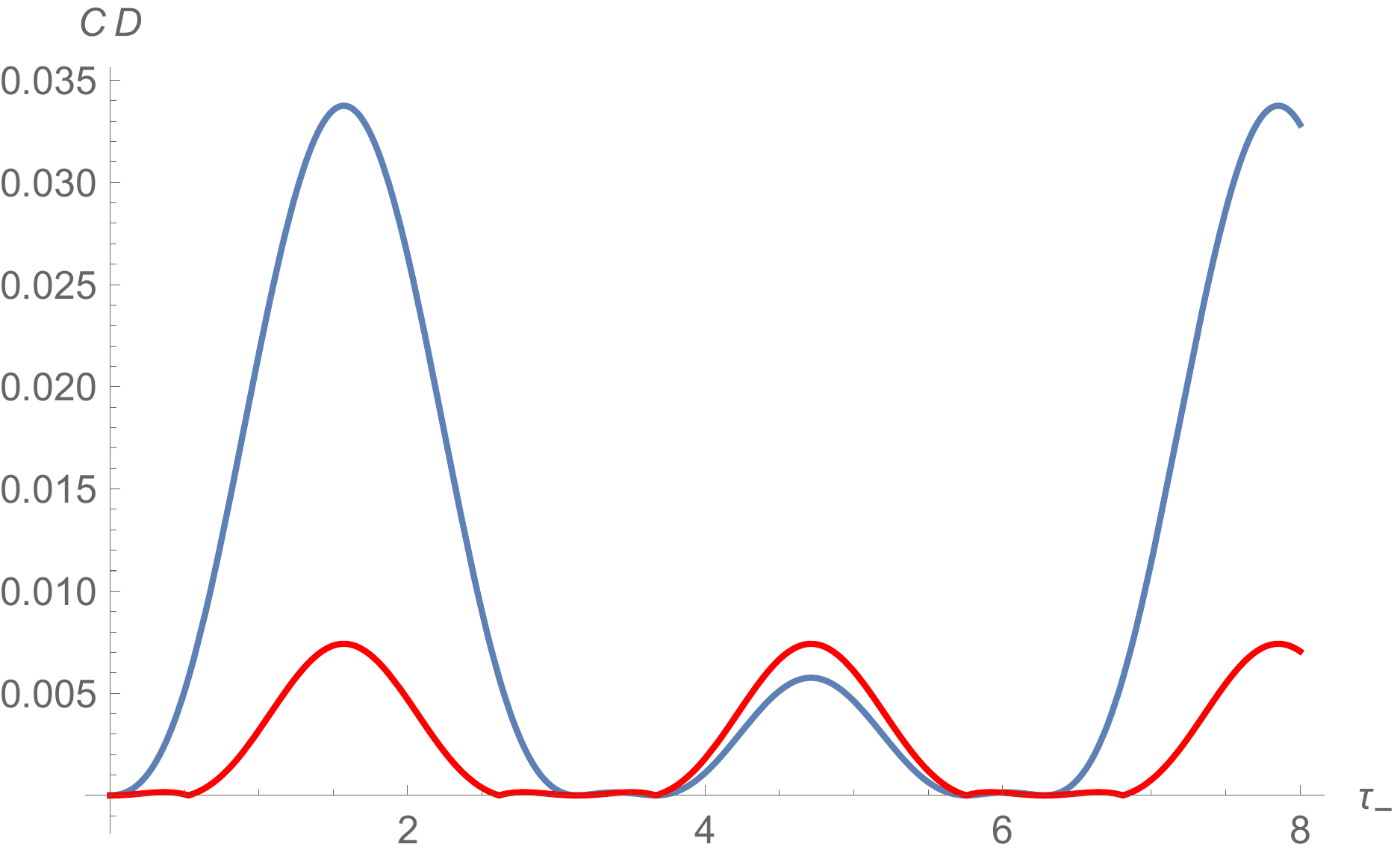}
    \includegraphics{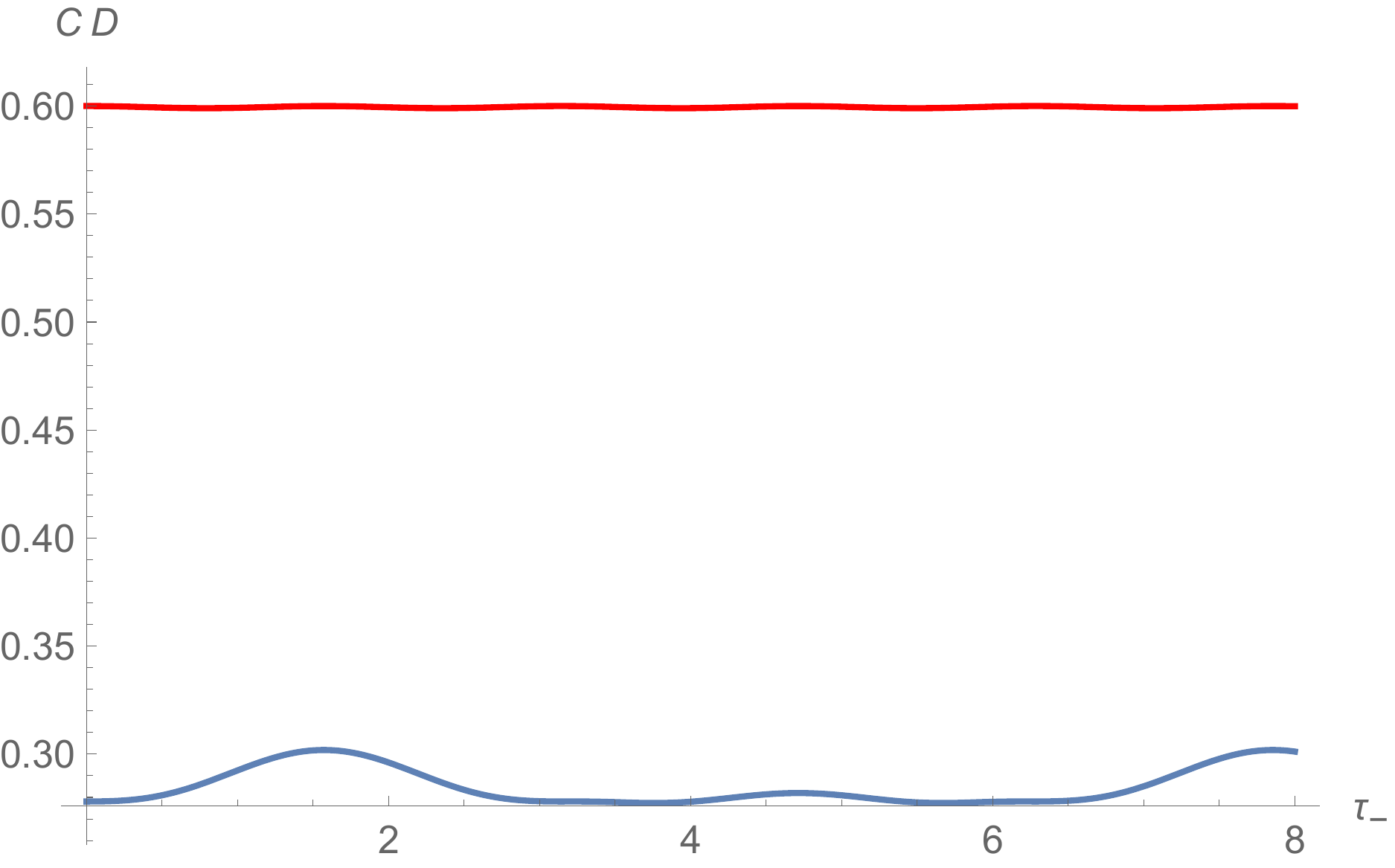}
     \includegraphics{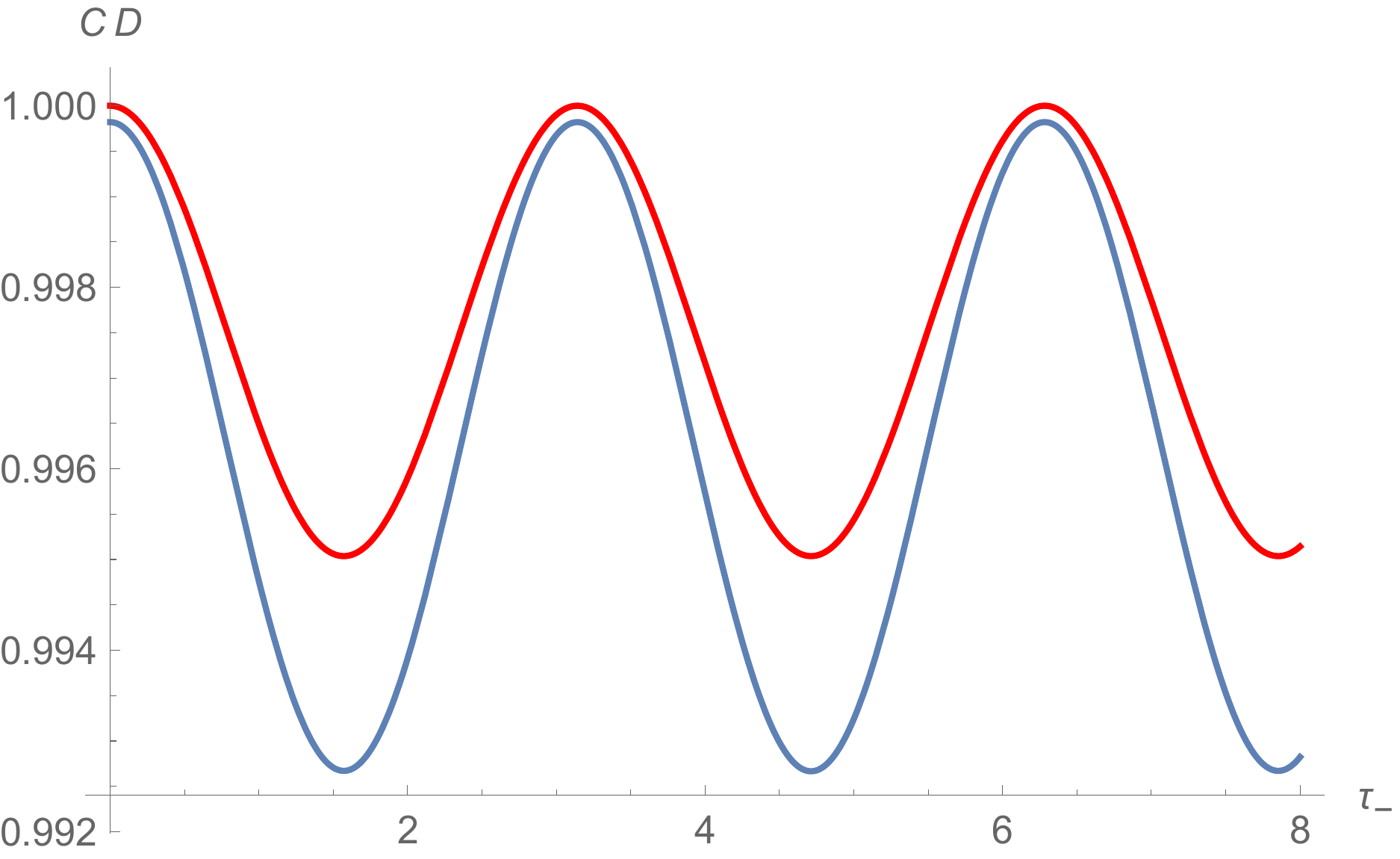}
}
\caption{Quantum concurrence $C$ (red) and quantum discord $D$ (blue) versus time $t$, for various values of the mixing parameter: $p=0,0.1,0.3,0.5,0.8,1,$ respectively, for constant magnetic fields, with $\Omega_{+}=10c$, for initial state (\ref{mixt1})-up.}
\label{fig:4}       
\end{figure}

\begin{figure}
\resizebox{1\textwidth}{!}{%
  \includegraphics{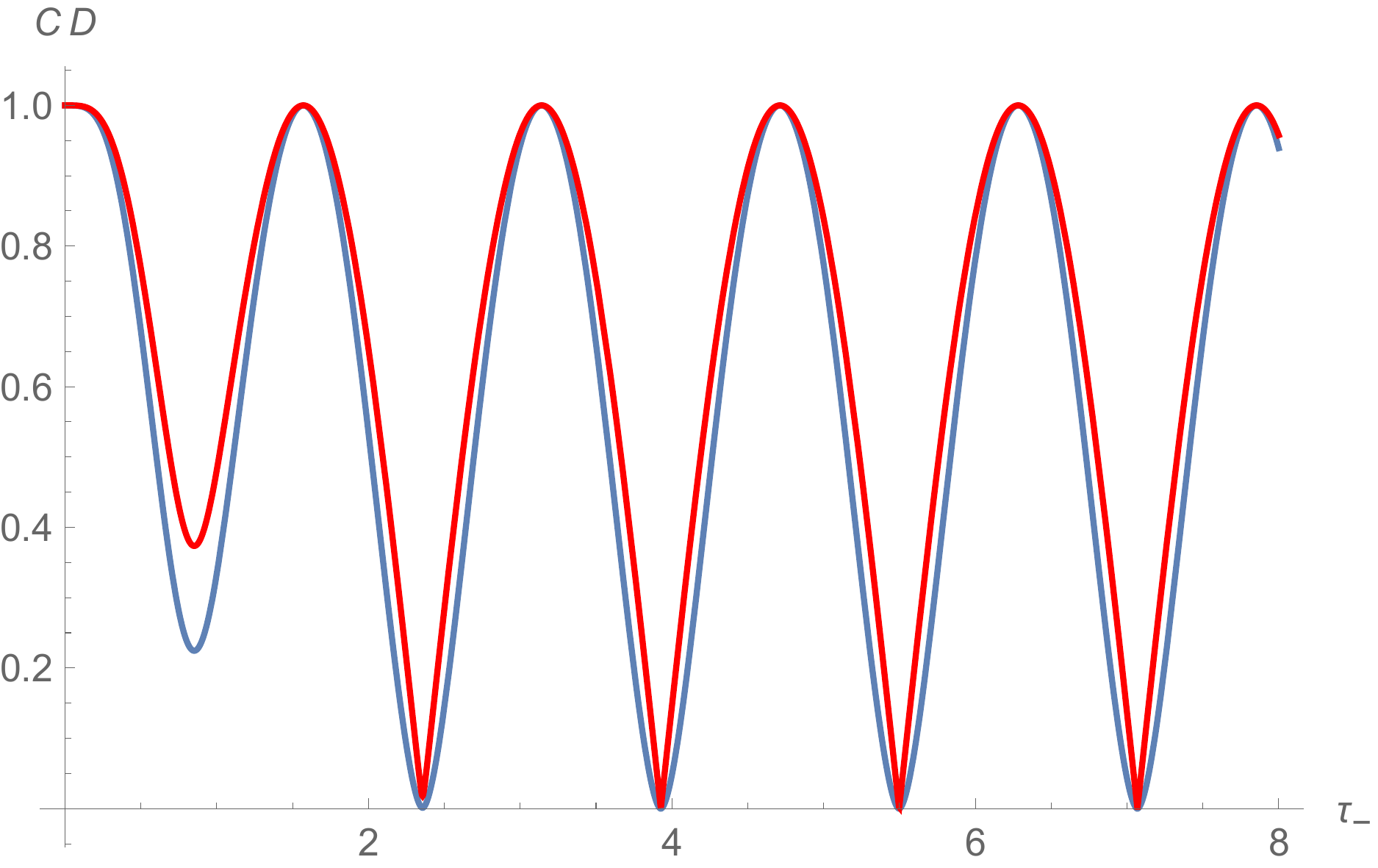}
   \includegraphics{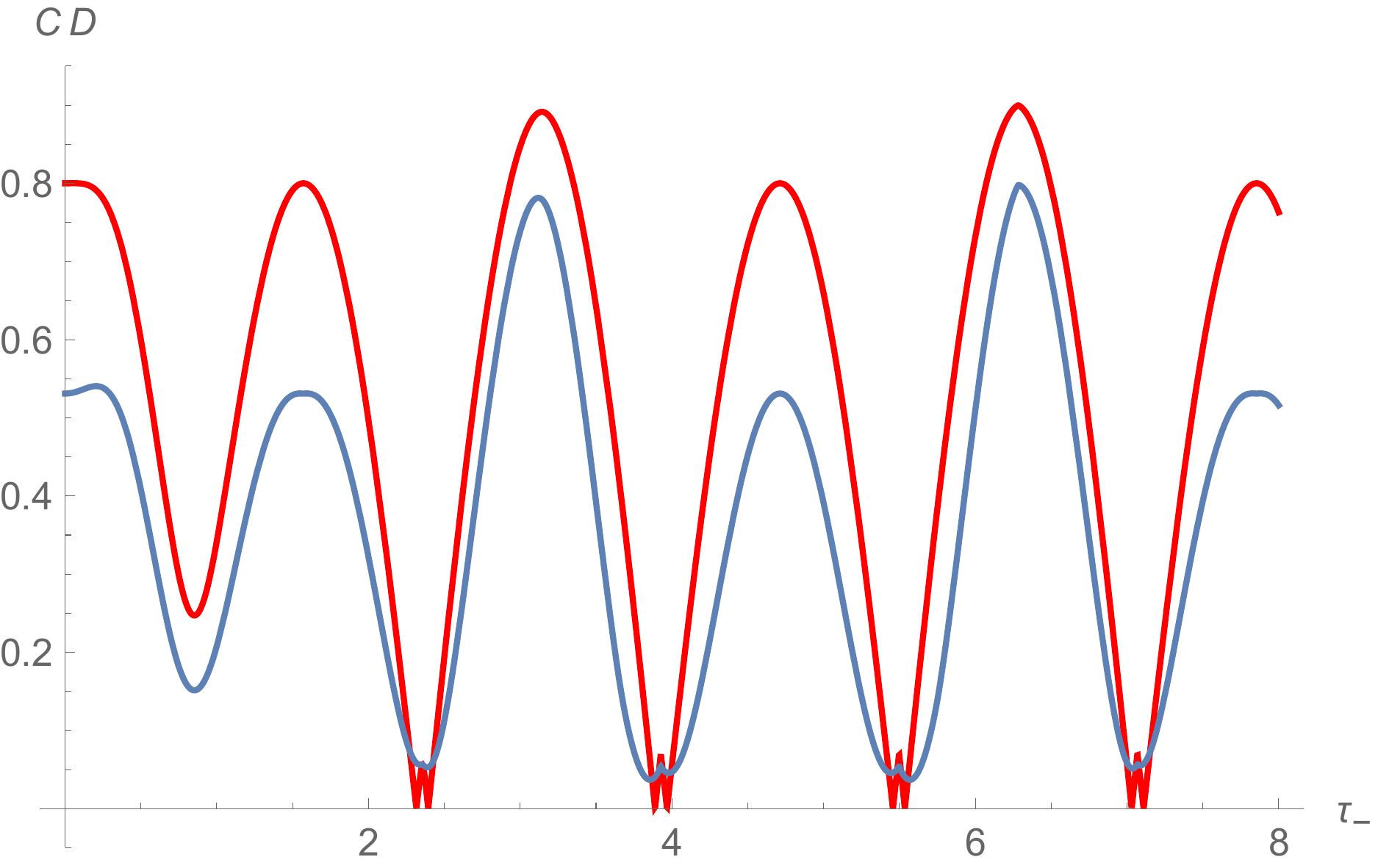}
    \includegraphics{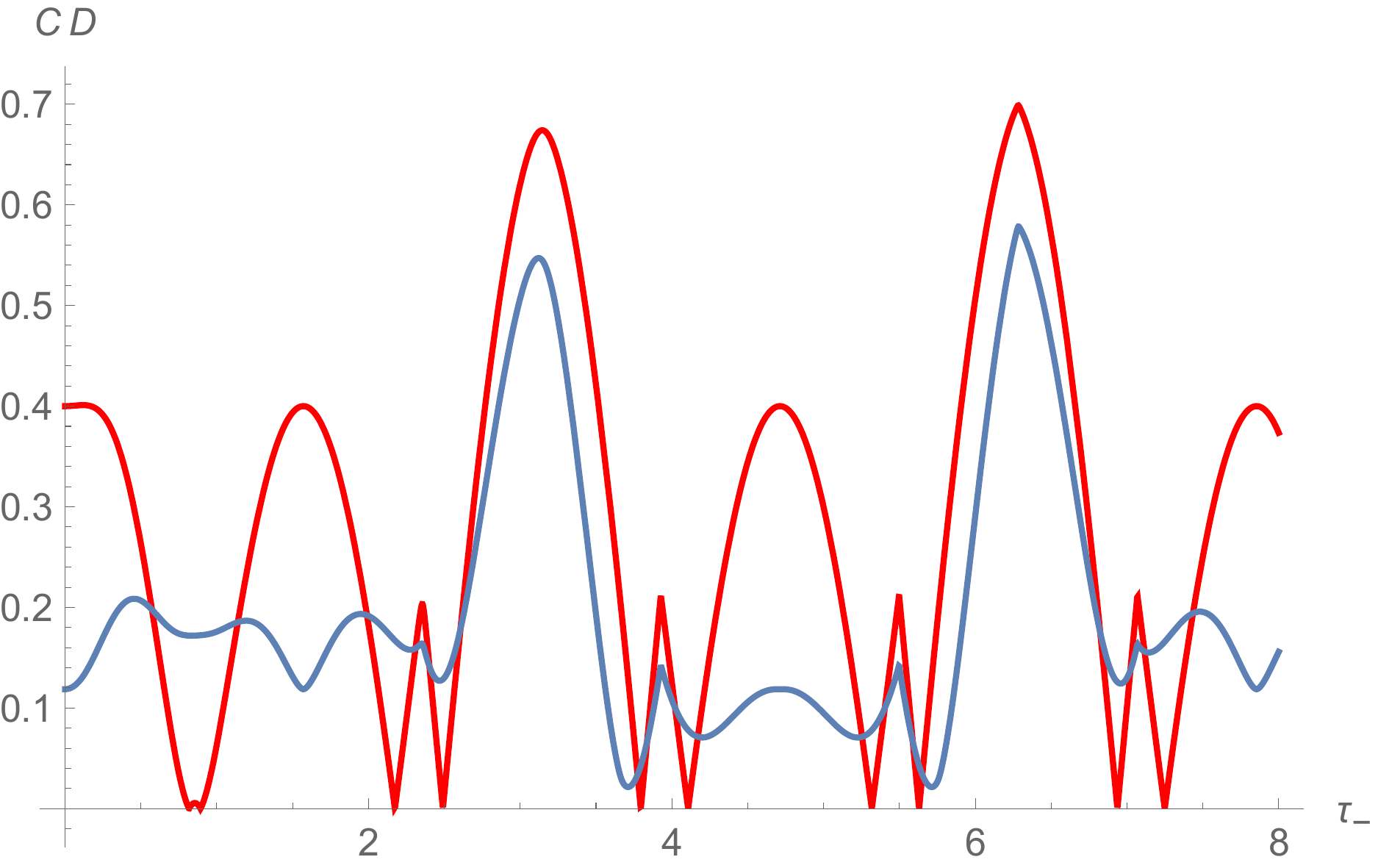}
    }
\resizebox{1\textwidth}{!}  {%
   \includegraphics{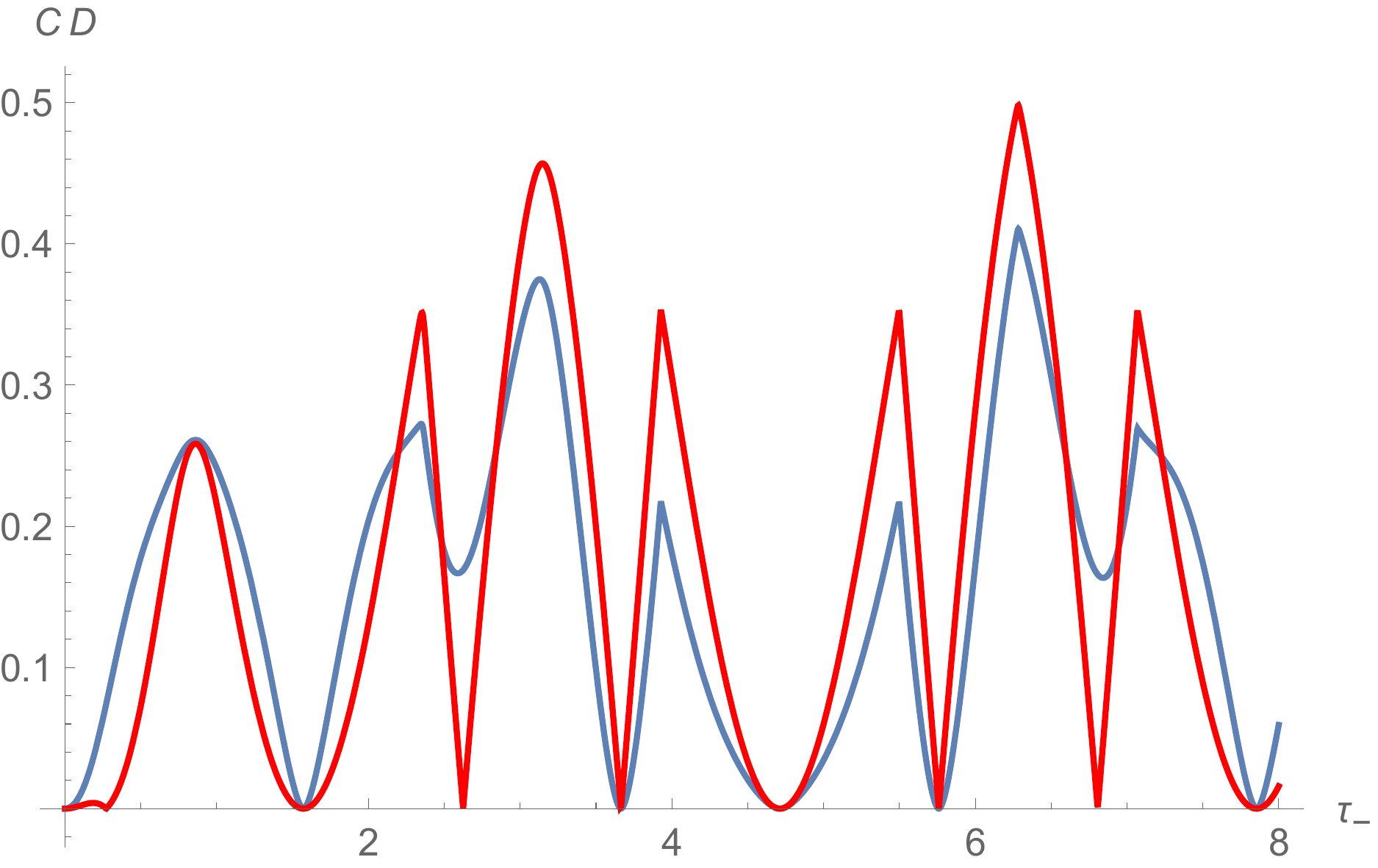}
    \includegraphics{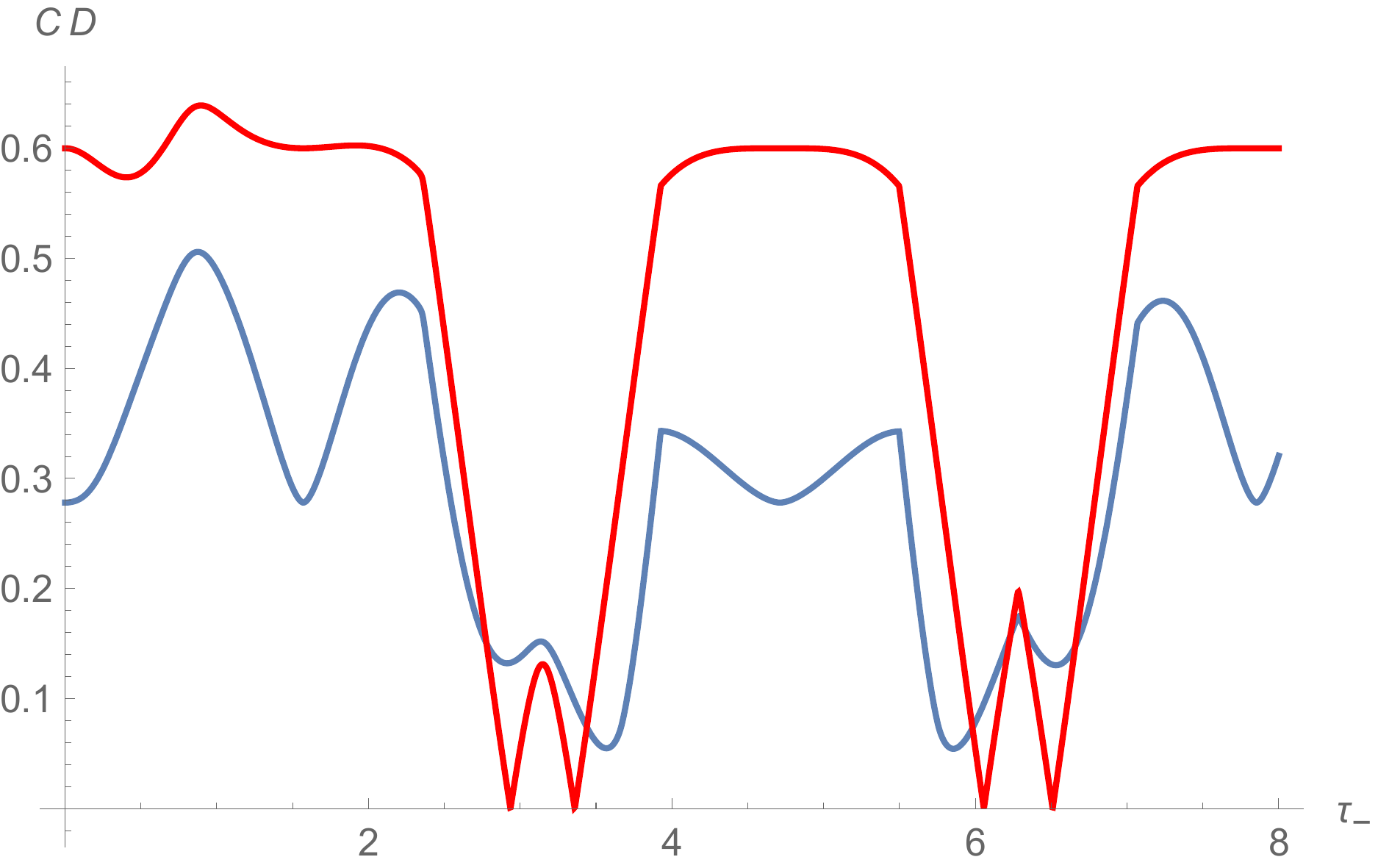}
     \includegraphics{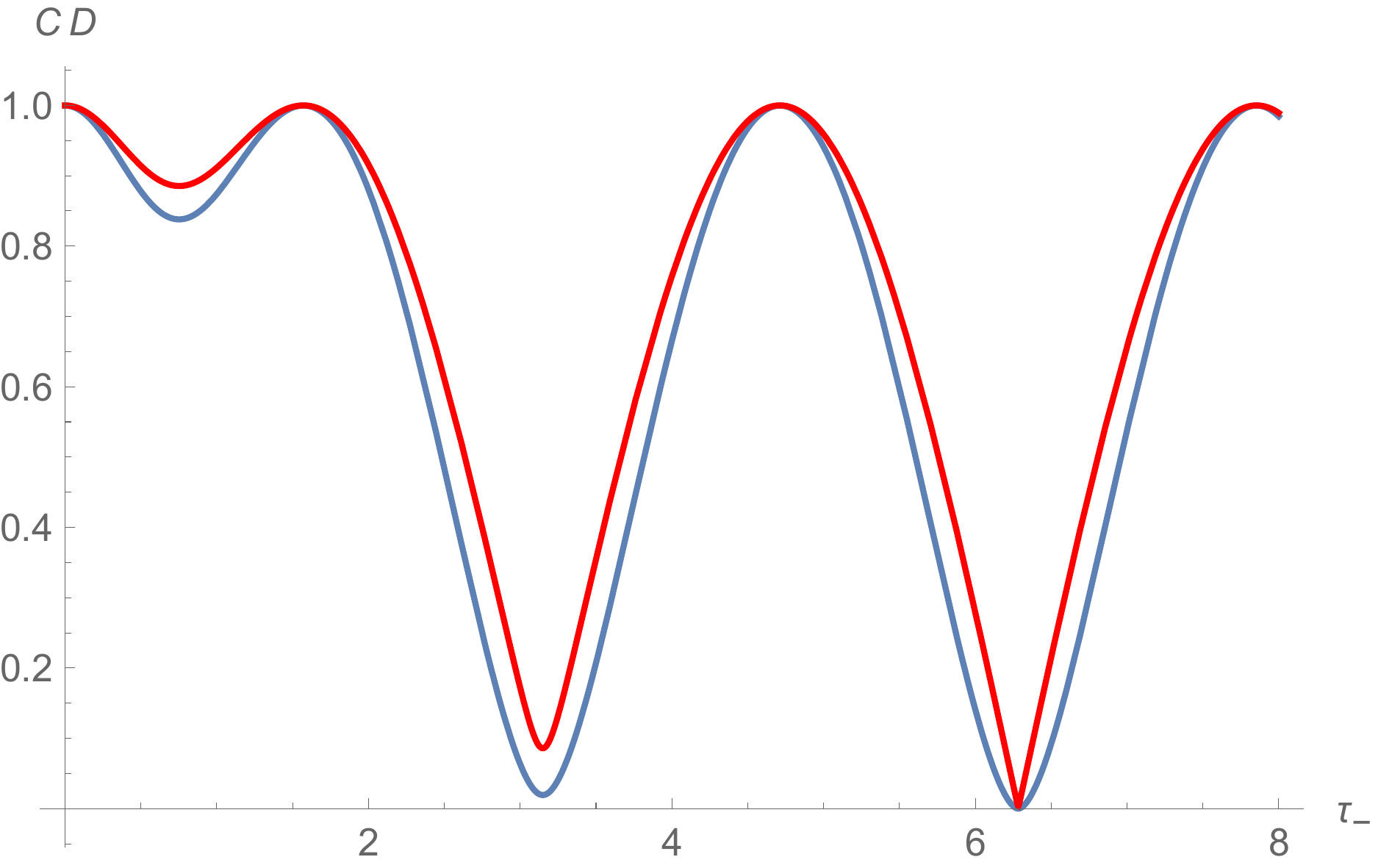}
}
\caption{Quantum concurrence $C$ (red) and quantum discord $D$ (blue) versus time $t$, for various values of the mixing parameter: $p=0,0.1,0.3,0.5,0.8,1,$ respectively, for the first model with (\ref{Quantities def ro 1 case}), for both initial states (\ref{mixt1})-up and (\ref{mixt3})-down.}
\label{fig:5}       
\end{figure}

\begin{figure}
\resizebox{1\textwidth}{!}{%
  \includegraphics{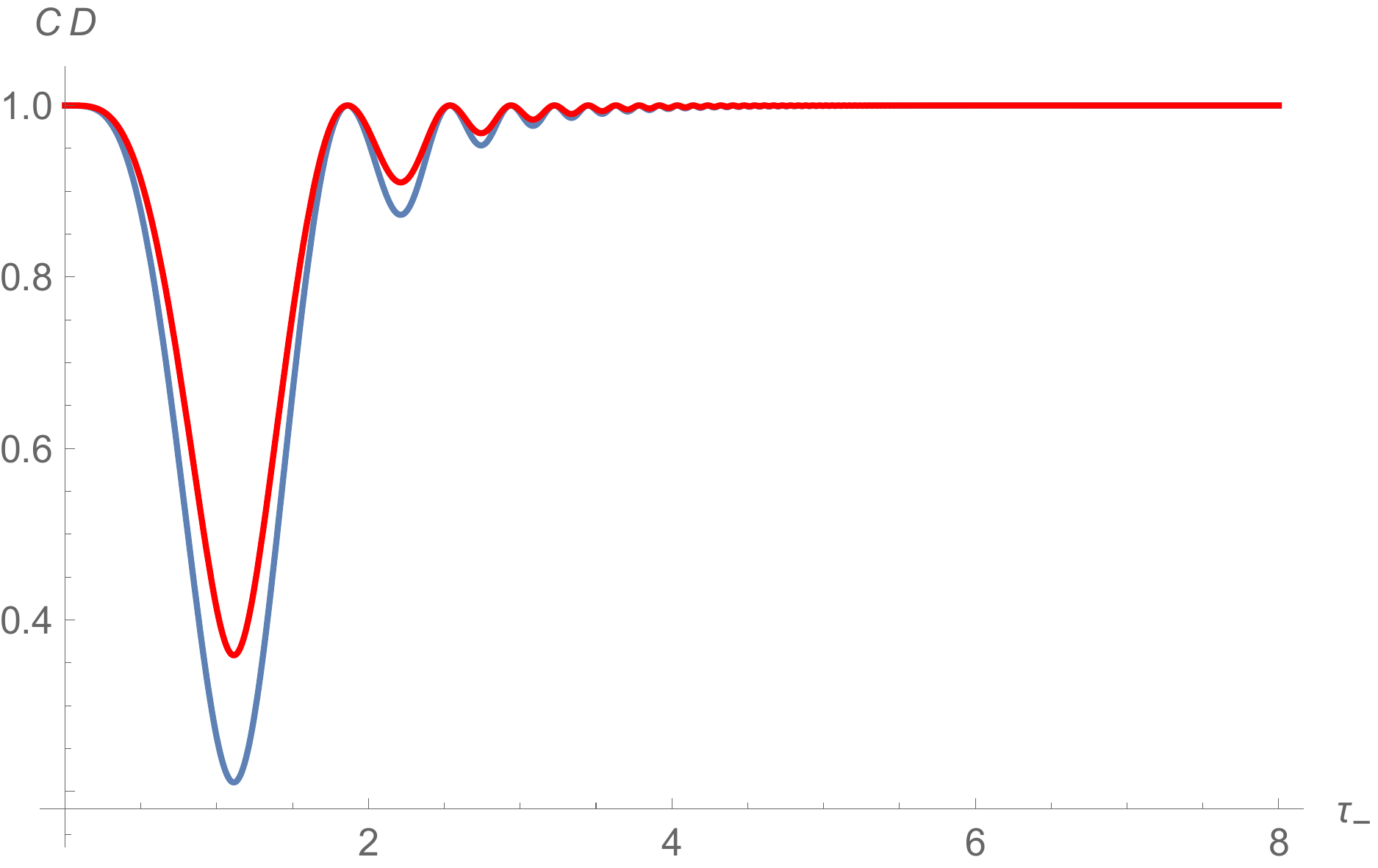}
   \includegraphics{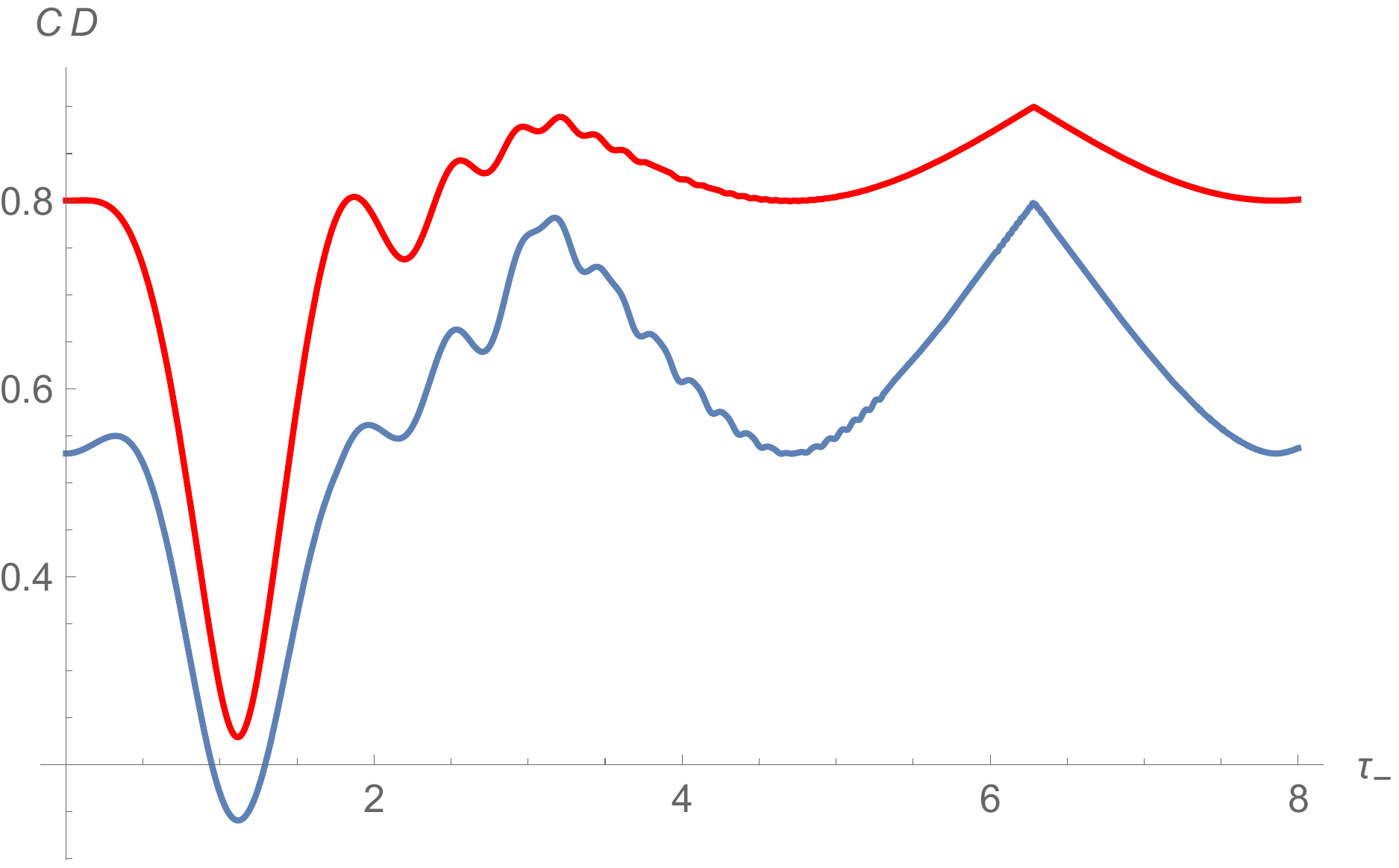}
    \includegraphics{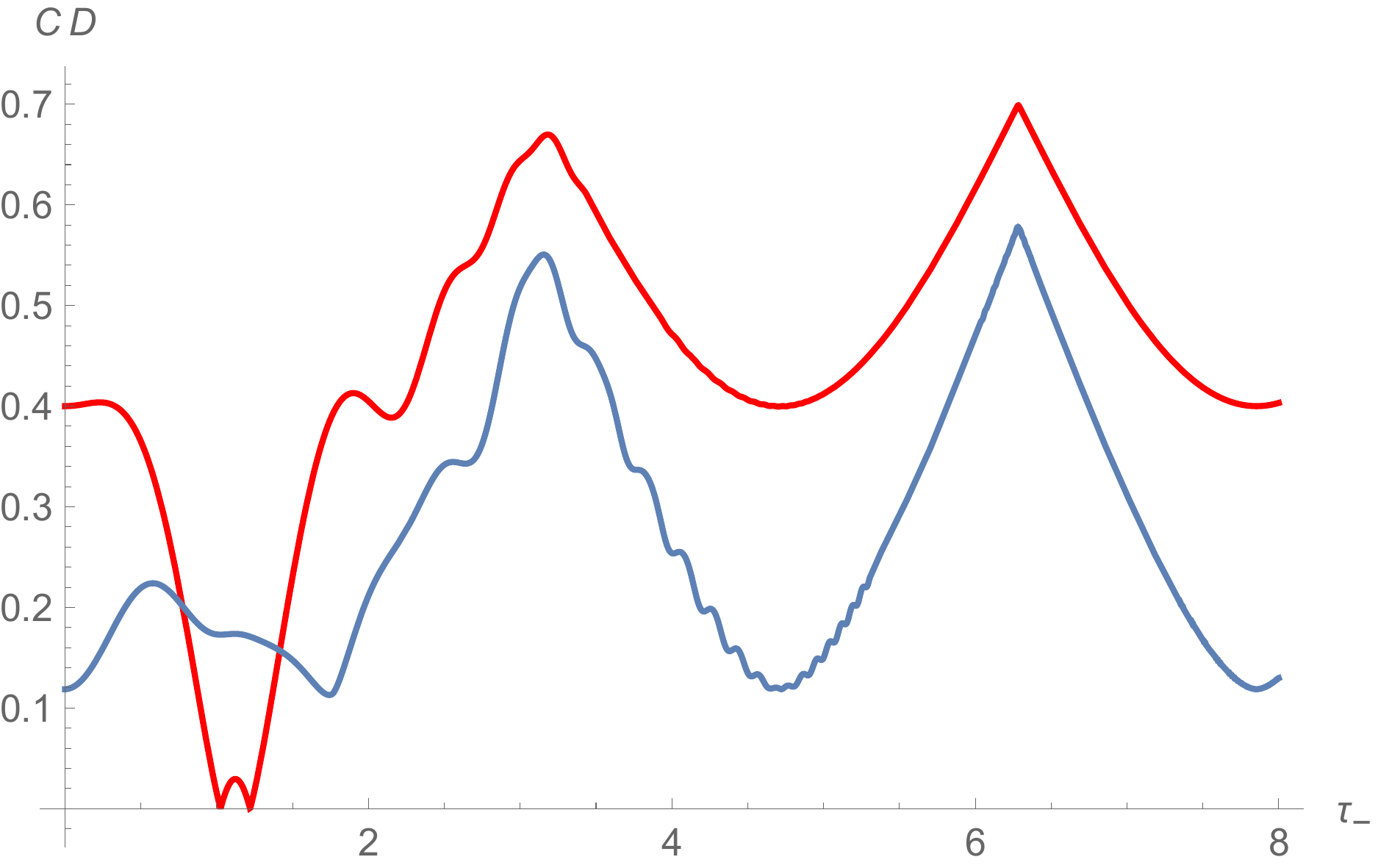}
   }
\resizebox{1\textwidth}{!}  {%
  \includegraphics{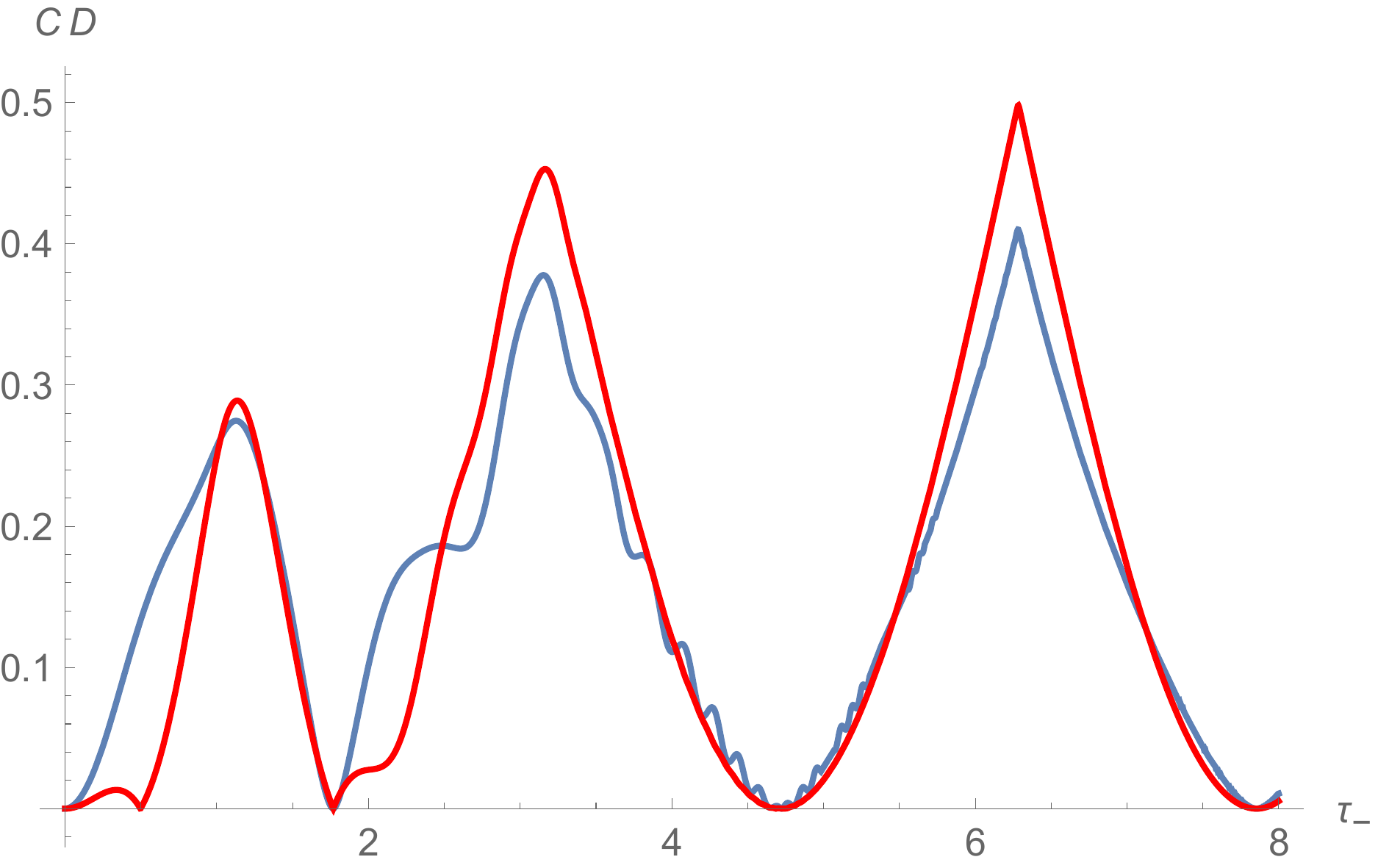}
   \includegraphics{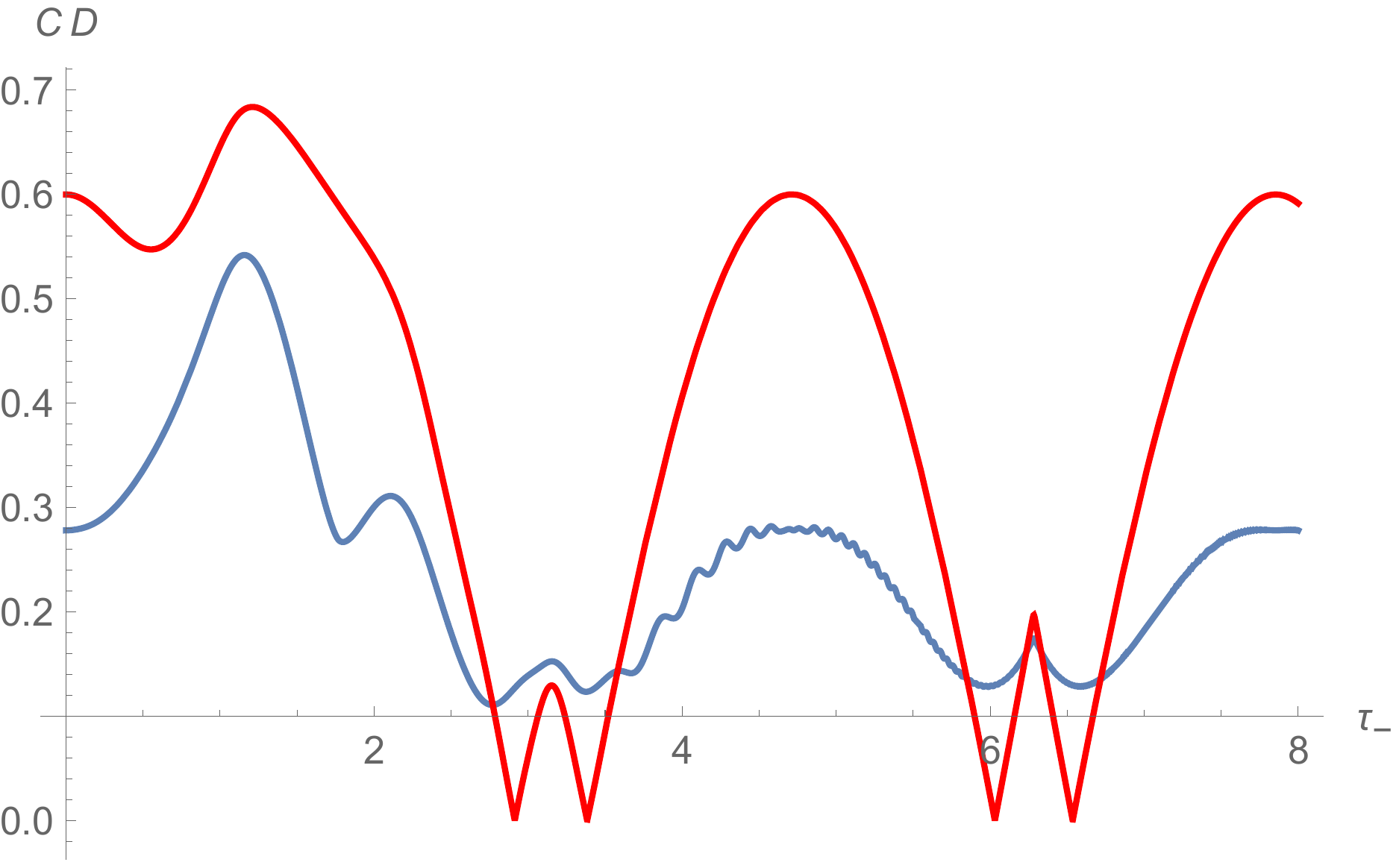}
    \includegraphics{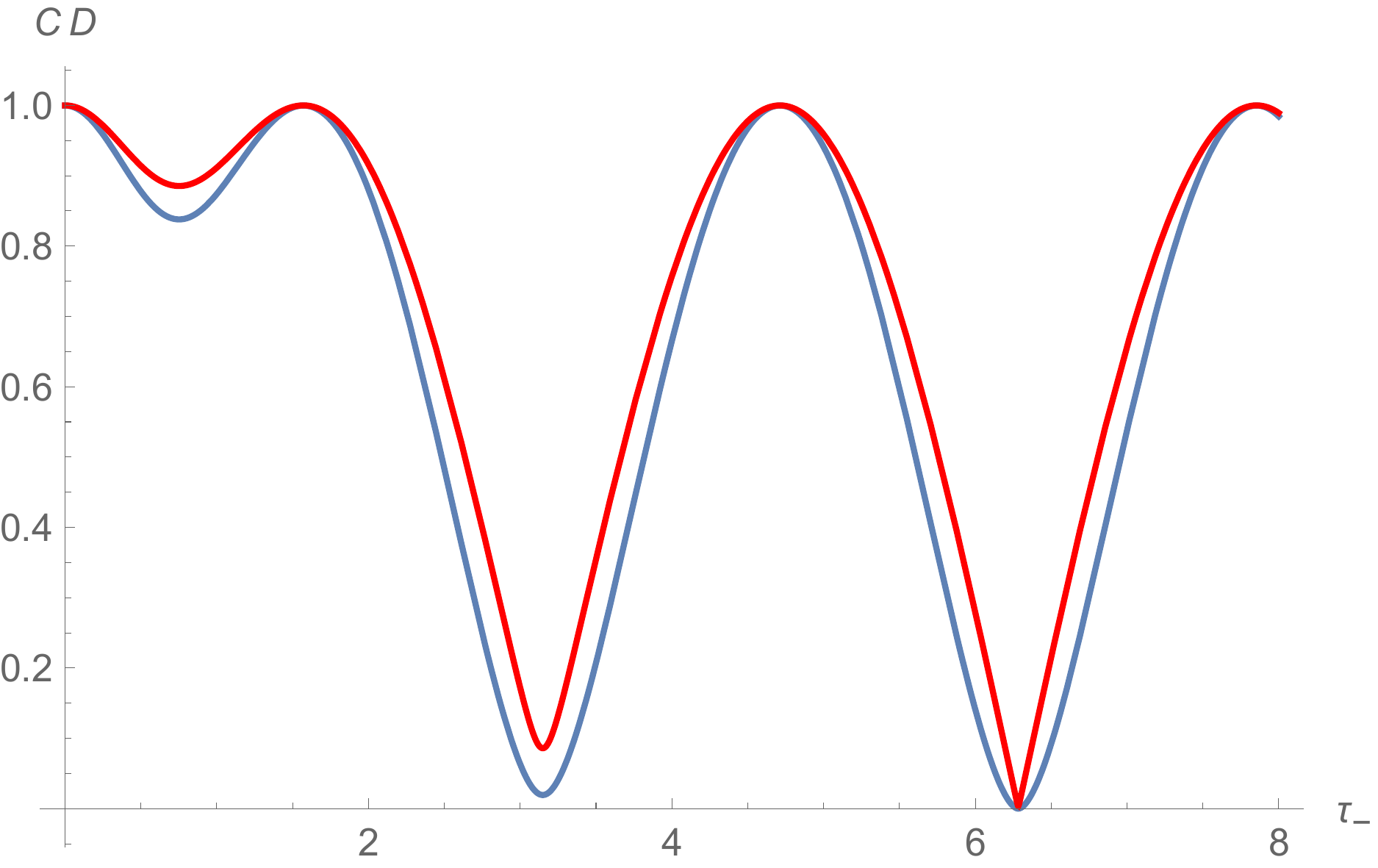}
}
\caption{Quantum concurrence $C$ (red) and quantum discord $D$ (blue) versus time $t$, for various values of the mixing parameter: $p=0,0.1,0.3,0.5,0.8,1,$ respectively, for the second model with (\ref{Quantities def ro 2 case}), for both initial states (\ref{mixt1})-up and (\ref{mixt3})-down.}
\label{fig:6}       
\end{figure}

\begin{figure}
\resizebox{1\textwidth}{!}{%
  \includegraphics{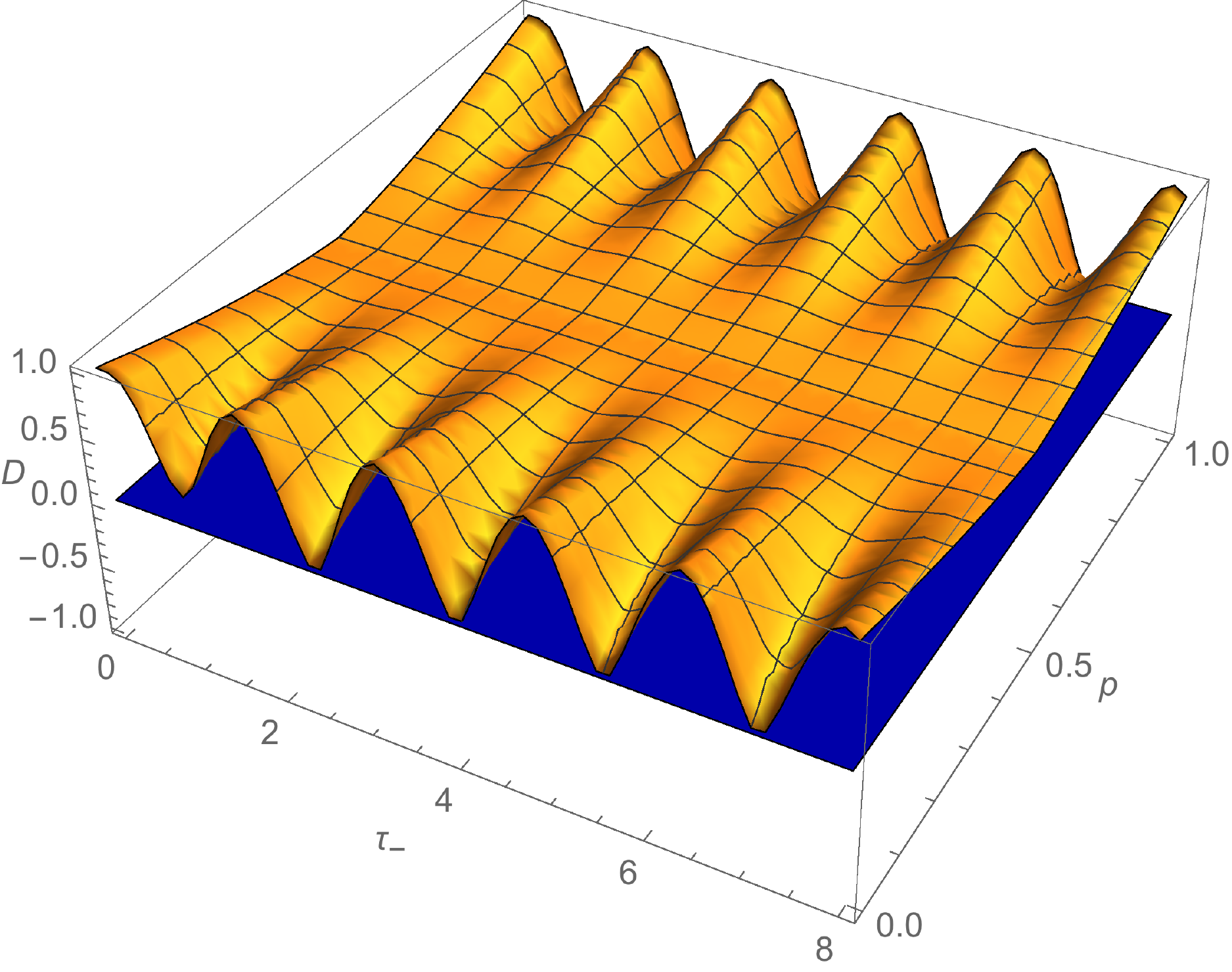}
    \includegraphics{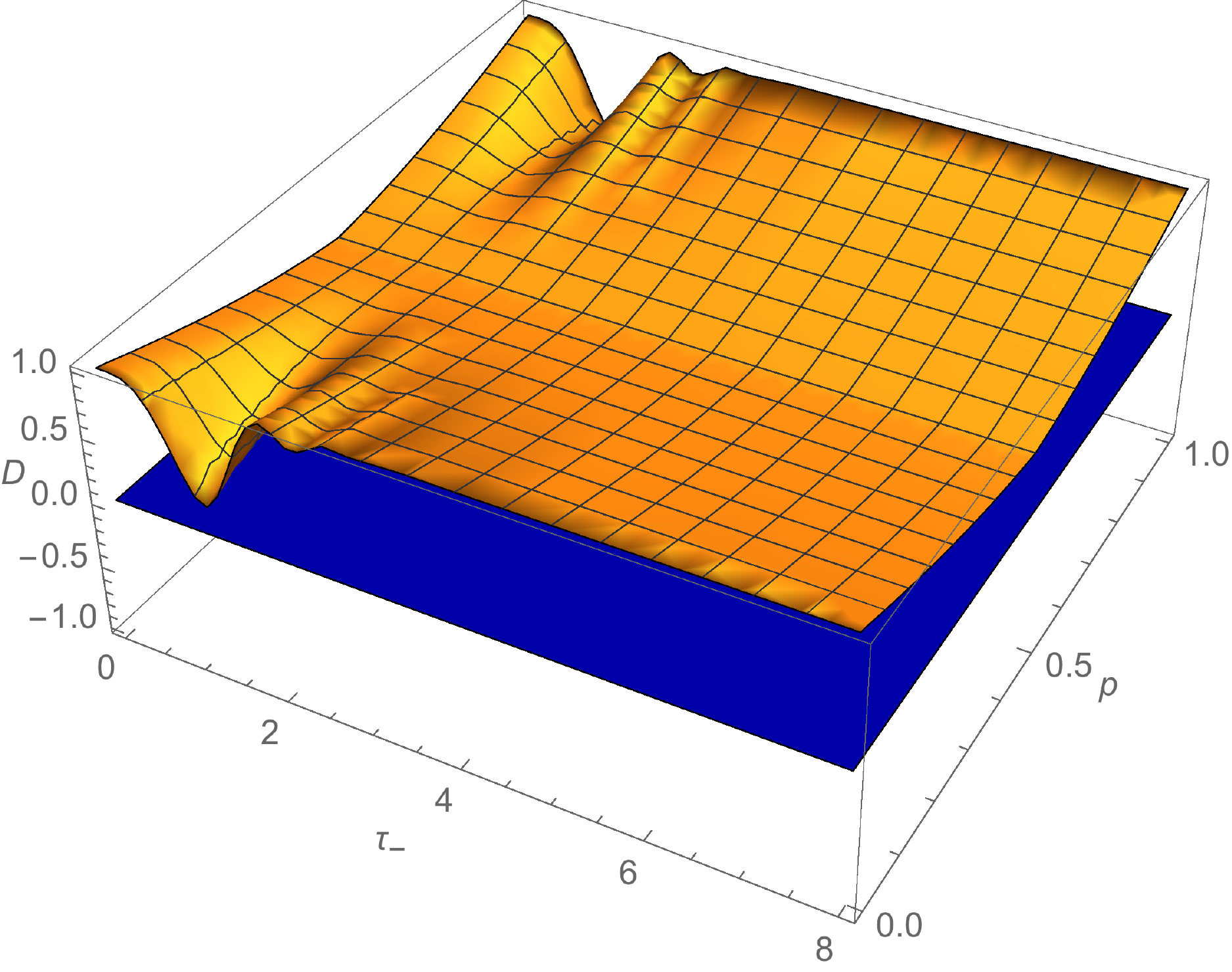}
     \includegraphics{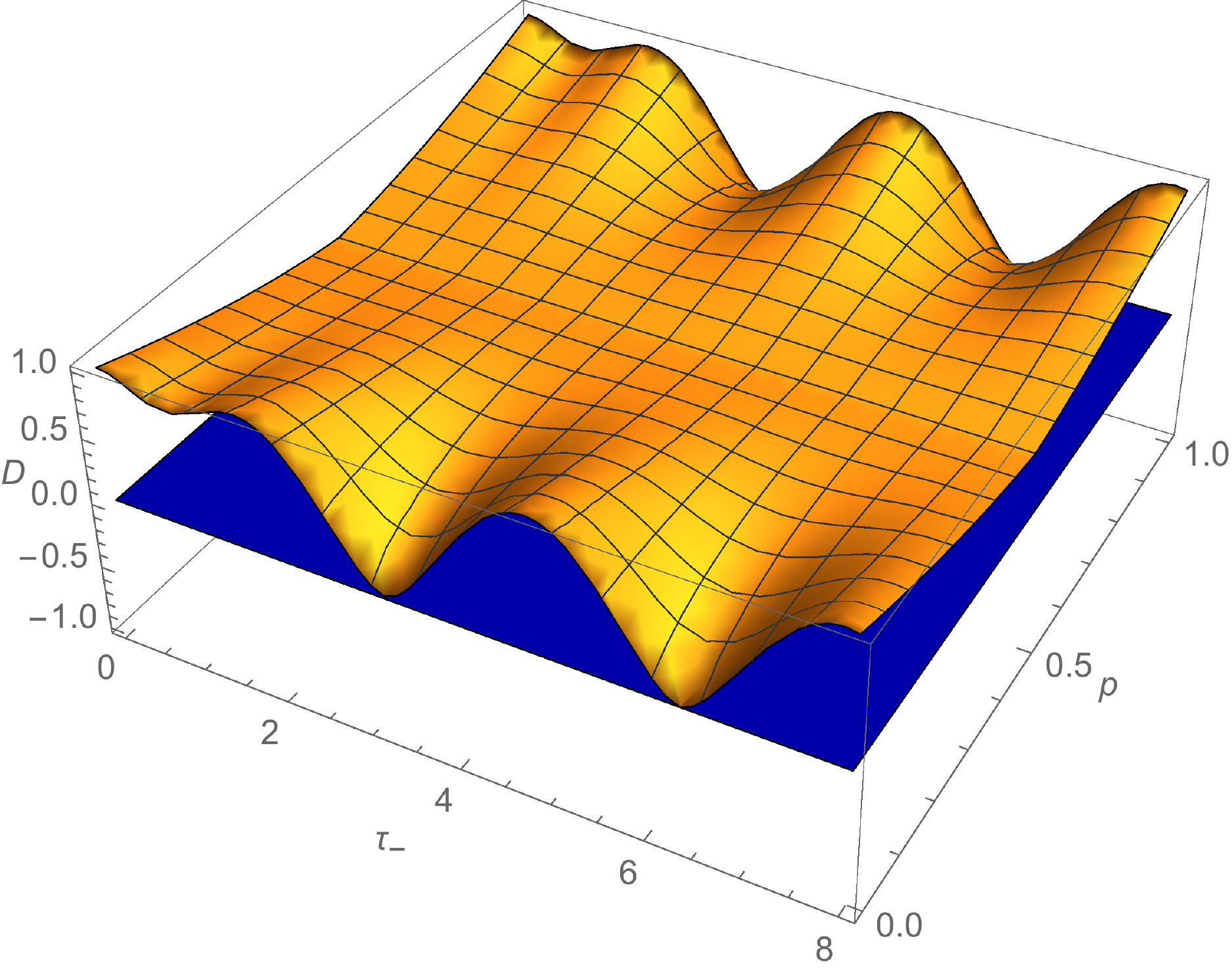}
   }
\resizebox{1\textwidth}{!}  {%
\includegraphics{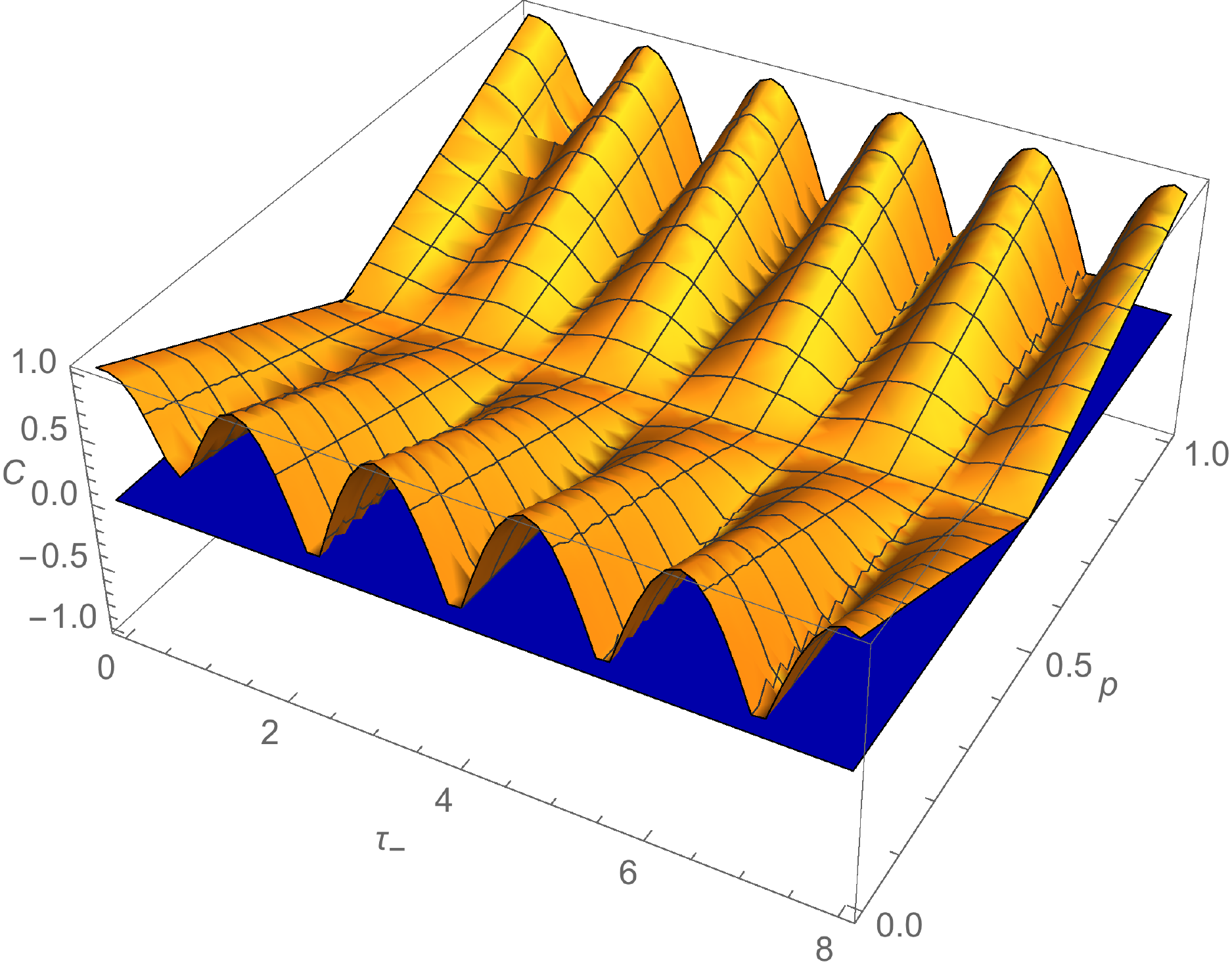}
     \includegraphics{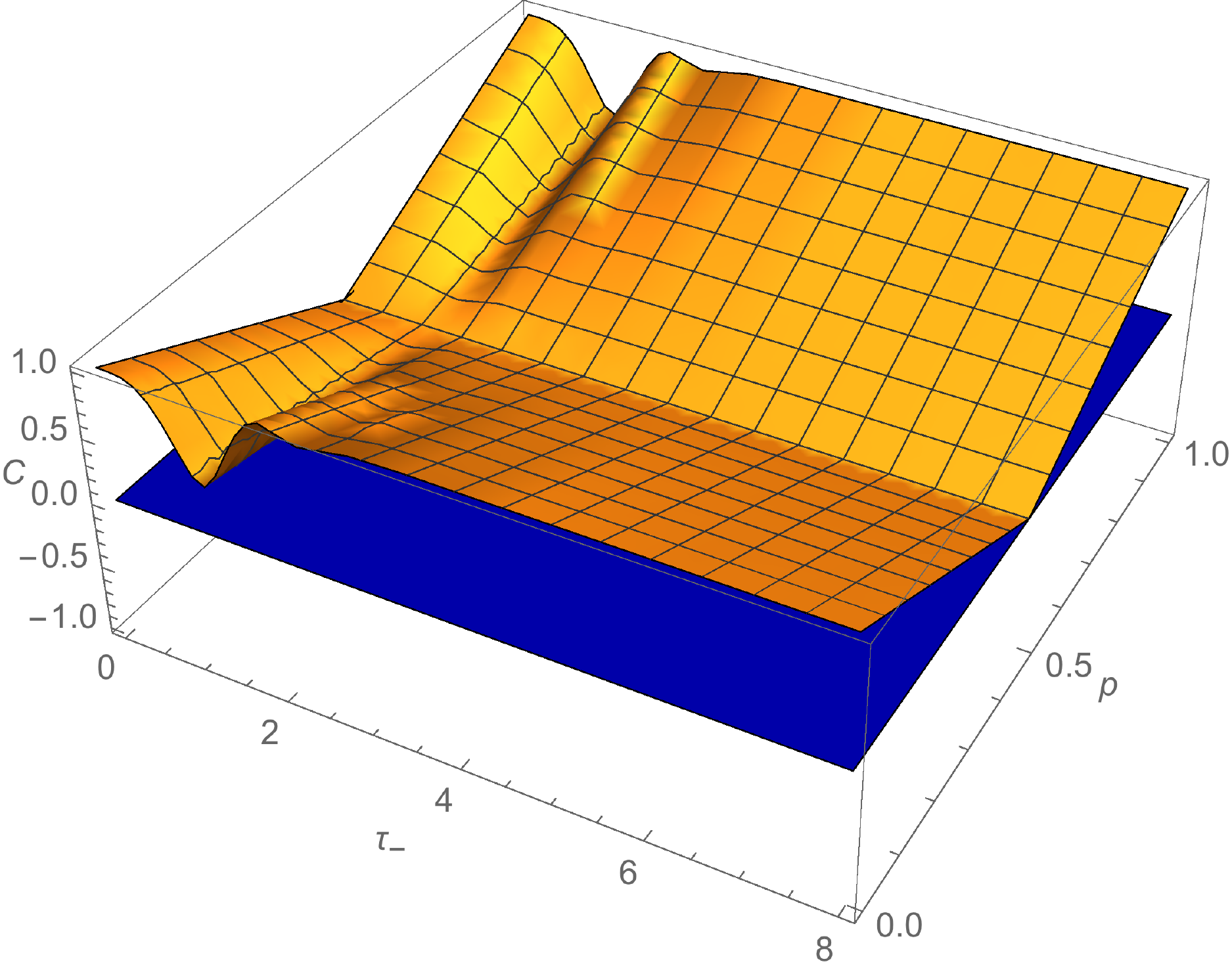}
     \includegraphics{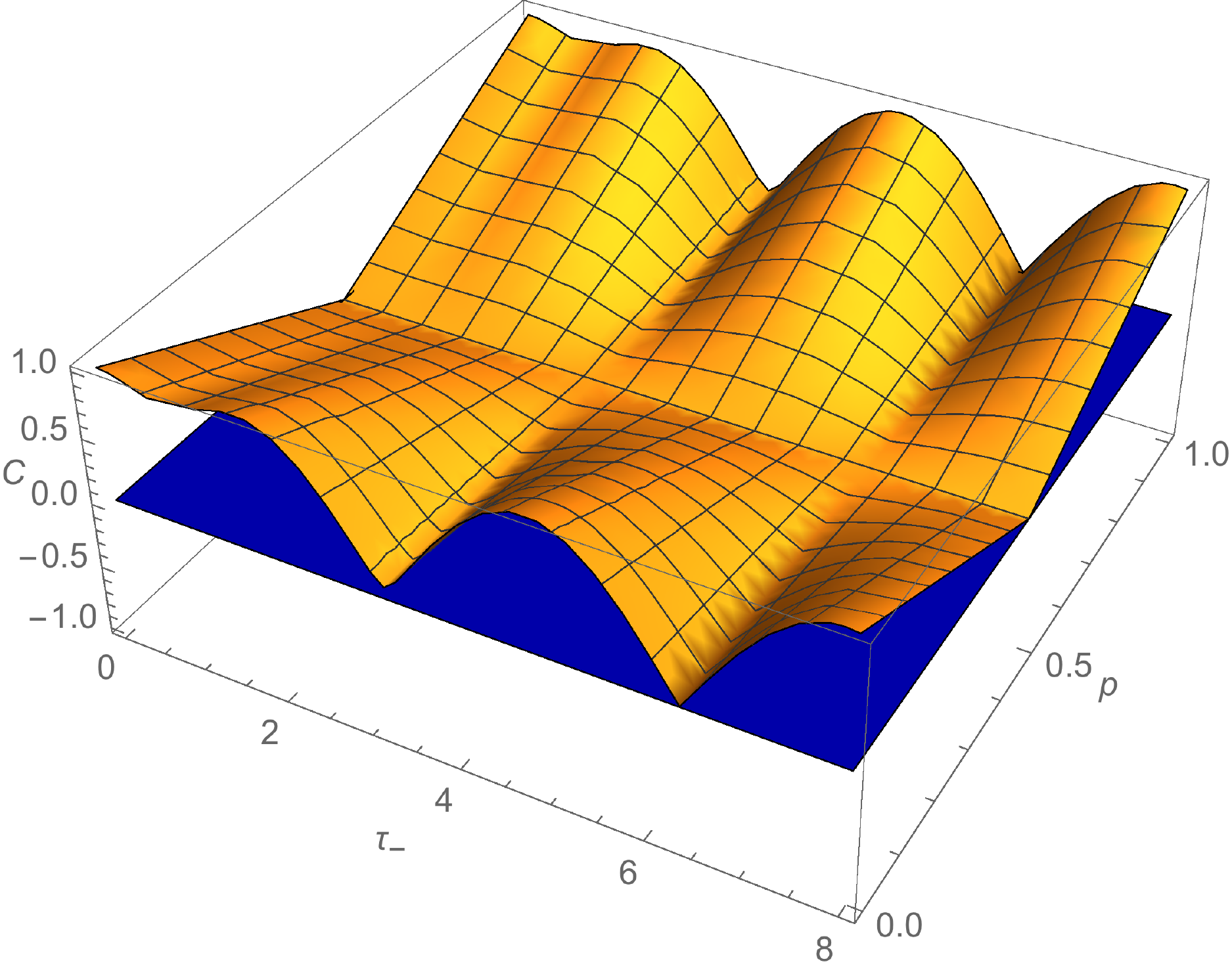}
}
\caption{
Quantum discord $D$ (up) and concurrence $C$ (down) versus time $t$ and mixing parameter $p$: for the first model with (\ref{Quantities def ro 1 case}), for the initial state (\ref{mixt6}) (left); for the second model with (\ref{Quantities def ro 2 case}), for the initial state (\ref{mixt6}) (middle); for both models with (\ref{Quantities def ro 1 case}), (\ref{Quantities def ro 2 case}), for the initial state (\ref{mixt5}) (right).
}
\label{fig:7}       
\end{figure}

In Figs. \ref{fig:1}, \ref{fig:2}, \ref{fig:3}, \ref{fig:4}, \ref{fig:5} and \ref{fig:6}, we have reported the time dependences of both the concurrence ($C$) and the quantum discord ($D$) for the state \eqref{mixt1}-up for different values of the mixing parameter $p$.
As expected, the dynamical behaviour of quantum discord is very similar to that of quantum entanglement, measured by the concurrence, and they evolve quite synchronously, especially for initial states having a small degree of mixedness ($p$ close to 0 or 1). They have an identical qualitative behaviour for initial pure states, but are different for the mixed states.

In Figs. \ref{fig:1}, \ref{fig:2}, \ref{fig:3} and \ref{fig:4}, $C$ and $QD$ are plotted for constant magnetic fields, precisely when they are both vanishing, very lower than, of the same order of and much larger than the interaction intensity, respectively.
In Fig. \ref{fig:3}, then, we may see the effects on $C$ and $QD$ purely due to the interaction between the two spins, according to our specific choice about the coupling parameters (see Appendix \ref{App A}).
Figs. \ref{fig:4}, \ref{fig:5} and \ref{fig:6}, instead, show us how the interplay between the interaction and the presence of static magnetic fields determines a different behaviour in time.
In the plots related to the values of $p=0$ for such figures, we may appreciate how the static magnetic field significantly changes the time-dependences of the two quantities under scrutiny.
In absence of magnetic fields we have, indeed, constant and maximum values for $C$ and $QD$, whereas when the magnetic field is present periodic oscillations occur.
Plots for $p=0.1,~0.3,~0.5,~0.8$ show instead a clear difference between the cases of low and high magnetic field.
For such states, indeed, we may identify qualitative different behaviours related to the intensity of the magnetic fields.
For high intensity magnetic fields it is worth noticing the appearance of plateaux in the time-behaviour of both $C$ and $QD$ in some cases for long interval of time.
In other case ($p=1/2$) such plateaux have a finite duration and makes legitimate to speak of sudden death and birth.
Moreover, we note phenomena of quasi-freezing of the quantum correlations when $p=0.1,~0.3,~0.8$.
Also the plots related to the value $p=1$ brings to light clearly effects of the different intensities of the magnetic fields on the time-dependence of $C$ and $QD$, even if in this case the same qualitative behaviour is maintained.
Precisely, we have a gradual squeezing of the curve towards the top value.
It is important to underline that the asymmetry in the time-behaviour between the states identified by $p=0$ and $p=1$ in Fig. \ref{fig:3} stems from the different Hamiltonians governing the two subdynamics which the two state belong to.
Precisely it is possible to verify that our specific choice of the coupling parameters determines such difference and in particular the fact that $C$ and $QD$ are constant when $p=0$.
It can be seen that if we choose $\gamma_{xx}=-\gamma_{yy}=2\gamma_{xy}=-2\gamma_{yx}$, we get an interchange of time-behaviour between the two states, with consequent constant quantum correlations for the state related to $p=1$.

In Figs. \ref{fig:5} and \ref{fig:6} we plot the time dependences of $C$ and $QD$ for the same states when the magnetic fields change in time accordingly to the two time-dependent scenarios defined in Eqs. (32)-(33) and (34)-(35).
We can see how a time-dependent magnetic field may deeply modify the time-behaviour of the quantum correlations arising between the two spins in the system.
This means that appropriately engineered magnetic fields may be used to control (generate, destroy, freeze) quantum correlations between the two subsystems as to manipulate them for several tasks.

In Fig. \ref{fig:7} finally, we report 3-D plots of $C$ and $QD$ as functions of time and $p$ for the other two initial conditions in Eqs. \eqref{mixt5} and \eqref{mixt6}.
A peculiar time-dependence may be noted for $p=1/2$; in this case, indeed, we have a constant value both for $C$ and $QD$.
Differently from the case of $p=0$ in Fig. \ref{fig:3}, such a behaviour is not due to our specific choice of the coupling parameters.
Rather, it is traceable back to the fact that the states in Eqs. \eqref{mixt5} and \eqref{mixt6} for $p=1/2$ do not evolve in time since they commute with $H$ at any time, as it is easy to see by direct inspection.

\section{Summary and conclusions}
\label{sec:5}

We analyzed the dynamics of the quantum discord of two interacting spin-1/2's described by the Hamiltonian model studied in detail in Ref. \cite{grimess}.
Such a model possesses a $C_2$-symmetry with respect to the quantization axis $z$ and this fact gives us the possibility of solving exactly the dynamics of the system by reducing the problem into two independent problems of single spin-1/2.
Such a reduction is valid also when the Hamiltonian parameters are time-dependent, allowing the study of the dynamics of the two interacting spins when they are subjected to specific time-dependent magnetic fields.
The hyperbolic secant time-dependence involved in our proposals was introduced by Rosen and Zener in the early 1930s investigating the quantum dynamics of a single spin \cite{Rosen}.
Since this sech pulse is experimentally realizable \cite{Economu,Greilich,Poem}, it appears in other generalized spin models of physical interest from both a theoretical and applicative point of view \cite{Hioe,Kyoseva,Vitanov}.
In view of our Hamiltonian model, it is worth noting that the Scanning Tunneling Microscopy (STM) allows the local application of magnetic fields on a single qubit while it interacts with other ones (e.g. in a spin chain)  \cite{Khajetoorians,Yan,Bryant,Tao,Lutz,Wieser,Sivkov}.
Such local fields are effective magnetic fields stemming from the tunable exchange interaction between the target spin we wish to address and the spin present on the STM tip \cite{Khajetoorians,Yan,Bryant,Tao,Lutz,Wieser,Sivkov}.
Thanks to the possibility of varying the distance between the tip spin and the one in the chain, effective time-dependent magnetic fields may be generated \cite{Wieser}.

The fundamental symmetry of the model is at the basis also of the other important property possessed by such a system consisting in the fact that an initial X-state maintains this structure at any time.
We know that quantum discord is a very difficult quantity to be calculated, but for the specific class of X-states we may take advantage of the analytic expression reported in Refs. \cite{ali,li}.
This enabled us to calculate the time evolution of the quantum discord for several mixed X-states consisting in convex combinations of Bell density matrices.
In particular, we have examined the cases of vanishing, very low and very high static magnetic fields with respect to the coupling constants strength.
In this way we could analyse the role of the magnetic fields in determining the occurrence of quantum correlations between the two spin-1/2's.
Furthermore, we have brought to light how specific time-dependences of the magnetic fields deeply modify the time-behaviour of the quantum discord, emphasizing how for such a model we may control (give rise, kill or freeze) the quantum correlations in time.

We also made a comparison of the behaviour of quantum discord and concurrence, as different measures of quantum correlations.
As expected, the dynamical behaviour of quantum discord and concurrence exhibit a similar behaviour when the system starts from pure states. The same comparison when the system starts from a mixed state, reported in this paper, makes evident the occurrence of remarkable differences analysed and discussed in Sec. 4.


Quantifying nonclassical correlations in physical systems enables a deeper understanding of genuine quantum behavior \cite{Gu}.
Over the last years, several methods and protocols have been developed in order to grasp experimentally signatures of such correlations \cite{Soares,Silva,USingh,Lekshmi}.
In Ref. \cite{Lanyon} an experimental algorithm allowing the tomographic reconstruction of the density matrix of a two-qubit system has been proposed as basis to evaluate quantum discord.
It is worth noticing how the diagonal Bell states, studied in this paper, play a prominent role for such kind of investigations thanks to their properties and their robustness exhibited also in open quantum systems immersed in dephasing environment \cite{Xu,jqli,wang,Singh}.

\section*{Acknowledgements}

The work of I.G. was supported by the funding agency CNCS-UEFISCDI of the Romanian Ministry of Research and Innovation through grant PN-III- P4-ID-PCE-2016-0794.
R.G. acknowledges economical support by research funds difc 3100050001d08+, University of Palermo, in memory of Francesca Palumbo.

\appendix

\section{Exactly solvable time-dependent scenarios}\label{App A}

In Ref. \cite{grimess} the authors shows that if we choose the two magnetic fields acting upon the two spin-1/2's as follows
\begin{equation}
\hbar \omega_{1/2}(t) = \dfrac{|\Gamma_+|}{\cosh(2\tau_+)} \pm \dfrac{|\Gamma_-|}{\cosh(2\tau_-)}
\end{equation}
the solutions for the entries of the time evolution operator \eqref{Total time ev op} are
\begin{equation} \label{a and b case 1}
\begin{aligned}
&|a_+(t)| = \sqrt{\dfrac{ \cosh(2\tau_+) + 1 }{2 \cosh(2\tau_+)}}, \qquad
 |b_+(t)| = \sqrt{\dfrac{ \cosh(2\tau_+) - 1 }{2 \cosh(2\tau_+)}}, \\
&\phi_{a}^+(t) = - \arctan [ \tanh ( \tau_+ ) ] - \tau_+ \qquad
 \phi_b^+(t) = \phi_{\Gamma_+} - \arctan [ \tanh ( \tau_+ ) ] + \tau_+ - \frac{\pi}{2} \\
&|a_-(t)| = \sqrt{\dfrac{ \cosh(2\tau_-) + 1 }{2 \cosh(2\tau_-)}}, \qquad
 |b_-(t)| = \sqrt{\dfrac{ \cosh(2\tau_-) - 1 }{2 \cosh(2\tau_-)}}, \\
&\phi_{a}^-(t) = - \arctan [ \tanh ( \tau_- ) ] - \tau_- \qquad
 \phi_b^-(t) = \phi_{\Gamma_-} - \arctan [ \tanh ( \tau_- ) ] + \tau_- - \frac{\pi}{2}.
\end{aligned}
\end{equation}
If, instead, the two local magnetic fields change in time as
\begin{equation}
\hbar \omega_{1/2}(t) = \dfrac{|\Gamma_+|}{\cosh(2\tau_+)} \pm
{|\Gamma_-| \over 4} \biggl[ { 3 \over \cosh(\tau_-) } - \cosh(\tau_-) \biggr], \\
\end{equation}
the solutions, in this case, read
\begin{equation} \label{a and b case 2}
\begin{aligned}
&|a_+(t)| = \sqrt{\dfrac{ \cosh(2\tau_+) + 1 }{2 \cosh(2\tau_+)}}, \qquad
 |b_+(t)| = \sqrt{\dfrac{ \cosh(2\tau_+) - 1 }{2 \cosh(2\tau_+)}}, \\
&\phi_{a}^+(t) = - \arctan [ \tanh ( \tau_+ ) ] - \tau_+ \qquad
 \phi_b^+(t) = \phi_{\Gamma_+} - \arctan [ \tanh ( \tau_+ ) ] + \tau_+ - {\pi \over 2} \\
& |a_-(t)| = \dfrac{1}{\cosh(\tau_-)}, \qquad |b_-(t)| = \tanh(\tau_-) \\
& \phi_{a}^-(t)= - \arctan \Bigl[ \tanh \Bigl( {\tau_- \over 2} \Bigr) \Bigr] - {1 \over 2} \sinh (\tau_-), \qquad
  \phi_b^-(t) = \phi_{\Gamma_-} - \arctan \Bigl[ \tanh \Bigl( {\tau_- \over 2} \Bigr) \Bigr] + {1 \over 2} \sinh (\tau_-) - {\pi \over 2}.
\end{aligned}
\end{equation}
In the previous expressions we put
\begin{equation}
\tau_\pm={|\Gamma_\pm| \over \hbar}t, \qquad |\Gamma_\pm|=\sqrt{(\gamma_{xx} \mp \gamma_{yy})^2+(\pm \gamma_{xy} + \gamma_{yx})^2}, \qquad \phi_{\Gamma_\pm}=-\arctan \left[ {\pm \gamma_{xy} + \gamma_{yx} \over \gamma_{xx} \mp \gamma_{yy} } \right].
\end{equation}
Actually, other two possible exactly solvable scenarios may be constructed, namely when the magnetic fields are
\begin{equation}
\begin{aligned}
&\hbar \omega_{1/2}(t) ={|\Gamma_+| \over 4} \biggl[ { 3 \over \cosh(\tau_+) } - \cosh(\tau_+) \biggr] \pm
\dfrac{|\Gamma_-|}{\cosh(2\tau_-)}, \\
&\hbar \omega_{1/2}(t) ={|\Gamma_+| \over 4} \biggl[ { 3 \over \cosh(\tau_+) } - \cosh(\tau_+) \biggr] \pm
{|\Gamma_-| \over 4} \biggl[ { 3 \over \cosh(\tau_-) } - \cosh(\tau_-) \biggr].
\end{aligned}
\end{equation}
Moreover, further exactly solvable time-dependent scenarios for a single spin-1/2 may be found in Refs. \cite{MGMN,GdCNM}.

In the previous formulas, $\tau_+$ and $\tau_-$ are scaled dimensionless times acting as independent variables;
$\phi_{\Gamma_+}$ and $\phi_{\Gamma_-}$ are true parameters strictly related to the microscopic model.
In our calculations we consider the case analyzed in Ref. \cite{grimess}, namely $\gamma_{xx}=\gamma_{yy}=\beta\gamma_{xy}=\beta\gamma_{yx}=c$ with $\beta =2$; we get $|\Gamma_+|=c=|\Gamma_-|/2$ and $\phi_{\Gamma_+}=-\pi/2$, $\phi_{\Gamma_-}=0$.
Then $\tau_-=2\tau_+$.

If the magnetic fields acting upon the two spin 1/2's were constant, i.e. $\omega_{1/2}=const.$ and then $\Omega_\pm=const.$, for the entries of the time evolution operator we would have
\begin{equation}
\begin{aligned}
a_\pm(t) &= e^{\mp i\gamma_{zz}t/\hbar} \left[ \cos(\nu_\pm t/\hbar) - i {\Omega_\pm \over \nu_\pm} \sin(\nu_\pm t/\hbar) \right] \\
b_\pm &= -ie^{\mp i\gamma_{zz}t/\hbar} {\Gamma_\pm \over \nu_\pm} \sin(\nu_\pm t/\hbar),
\end{aligned}
\end{equation}
with $\nu_\pm \equiv \sqrt{\Omega_\pm^2+|\Gamma_\pm|^2}$.
For the constant magnetic field cases we consider:
\begin{itemize}
\item
$|\Gamma_+|=c=|\Gamma_-|/2$, $\phi_{\Gamma_+}=-\pi/2$, $\phi_{\Gamma_-}=0$;
\item
$\Omega_-=2\Omega_+, \quad \Omega_+=3c$;
\item
$\tau_\pm=\nu_\pm t/\hbar$.
\end{itemize}
So we have
\begin{itemize}
\item
$\nu_-=2\nu_+$ $\rightarrow$ $\tau_-=2\tau_+$, $\tau_+=\sqrt{10}ct/\hbar$;
\item
${\Gamma_+ \over \nu_+}={-i \over \sqrt{10}}$, ${\Gamma_- \over \nu_-}={1 \over \sqrt{10}}$;
\item
${\Omega_+ \over \nu_+}={\Omega_- \over \nu_-}={1 \over \sqrt{1,1}}$.
\end{itemize}

\section{X-states and their evolution}\label{App B}

The general and formal expressions of the entries of $\rho(t)=U(t)\rho_XU^\dagger(t)$ may be written as follows
\begin{equation}
\begin{aligned}
&\rho_{11}(t)= |a_+|^2\rho_{11}+|b_+|^2\rho_{44}+2\text{Re}[a_+b_+^*\rho_{14}] \\
&\rho_{14}(t)= a_+^2\rho_{14}-b_+^2\rho_{41}-a_+b_+(\rho_{11}-\rho_{44})=\rho_{41}^*(t) \\
&\rho_{22}(t)= |a_-|^2\rho_{22}+|b_-|^2\rho_{33}+2\text{Re}[a_-b_-^*\rho_{23}] \\
&\rho_{23}(t)= a_-^2\rho_{23}-b_-^2\rho_{32}-a_-b_-(\rho_{22}-\rho_{33})=\rho_{32}^*(t) \\
&\rho_{33}(t)= |b_-|^2\rho_{22}+|a_-|^2\rho_{33}-2\text{Re}[a_-b_-^*\rho_{23}] \\
&\rho_{44}(t)= |b_+|^2\rho_{11}+|a_+|^2\rho_{44}-2\text{Re}[a_+b_+^*\rho_{14}],\label{elem}
\end{aligned}
\end{equation}
with all the other entries equal to 0.
The $X$-state $\rho(t)$ of Eq. \eqref{elem} has complex entries.
By performing the local unitary operation $U_A \otimes U_B$ described by Eq. \eqref{Loc Unit Transf}, $\rho(t)$ is transformed to its canonical form by replacing $\rho_{ij}$ by $|\rho_{ij}|$ for $i$ different from $j$.
By Eqs. \eqref{elem} it is straightforward to deduce the time-dependence of the five parameters in Eq. (\ref{Xstate}) in terms of the elements of the general density matrix in Eq. (\ref{X-state}), namely:
\begin{align}
  r(t) & = \rho_{11}(t)+\rho_{22}(t) -\rho_{33}(t)-\rho_{44}(t)
  ,\nonumber\\
  s(t) & = \rho_{11}(t)-\rho_{22}(t)+ \rho_{33}(t)-\rho_{44}(t)
  ,\nonumber\\
  c_{3}(t) & =\rho_{11}(t)-\rho_{22}(t)- \rho_{33}(t)+\rho_{44}(t) 
  ,\\
  c_{1}(t) & =2(|\rho_{23}(t)|+|\rho_{14}(t)|), \nonumber \\
  c_{2}(t) & =2(|\rho_{23}(t)|-|\rho_{14}(t)|). \nonumber
\end{align}

To get easily X-structured density matrices it is sufficient to consider convex combination of Bell states.
Let us consider, firstly, the Bell states as initial conditions, namely
\begin{equation}
\rho_0^\pm=\ket{\Phi^\pm}\bra{\Phi^\pm}={1 \over 2}
\left(
\begin{array}{cccc}
 1 & 0 & 0 & \pm 1 \\
 0 & 0 & 0 & 0 \\
 0 & 0 & 0 & 0 \\
 \pm 1 & 0 & 0 & 1 \\
\end{array}
\right),
\qquad
\tilde{\rho}_0^\pm=\ket{\Psi^\pm}\bra{\Psi^\pm}={1 \over 2}
\left(
\begin{array}{cccc}
 0 & 0 & 0 & 0 \\
 0 & 1 & \pm 1 & 0 \\
 0 & \pm 1 & 1 & 0 \\
 0 & 0 & 0 & 0 \\
\end{array}
\right).
\end{equation}
It is easy to see that $\rho^\pm(t)=U(t)\rho_0^\pm U^\dagger(t)$ keeps the same structure of $\rho_0^\pm$ at any time, so that the only entries changing in time are the following ones
\begin{equation}
\begin{aligned}
\rho_{11}^\pm(t)= {1 \over 2}\pm\text{Re}[a_+b_+^*] , \quad
\rho_{14}^\pm(t)= \pm{ a_+^2-b_+^2 \over 2 }, \quad
\rho_{44}^\pm(t)= {1 \over 2}\mp\text{Re}[a_+b_+^*].
\end{aligned}
\end{equation}
Analogously for $\tilde{\rho}^\pm(t)=U(t)\tilde{\rho}_0^\pm U^\dagger(t)$ we get
\begin{equation}
\begin{aligned}
\tilde{\rho}_{22}^\pm(t)= {1 \over 2}\pm\text{Re}[a_-b_-^*], \quad
\tilde{\rho}_{23}^\pm(t)= \pm{ a_-^2-b_-^2 \over 2 }, \quad
\tilde{\rho}_{33}^\pm(t)= {1 \over 2}\mp\text{Re}[a_-b_-^*].
\end{aligned}
\end{equation}

In the first case, the explicit expression of the density matrices for the four possible initial Bell states are dictated by the following quantities:
\begin{equation}\label{Quantities def ro 1 case}
\begin{aligned}
&\text{Re}[a_\pm b_\pm^*]={1 \over 2} \tanh(2\tau_\pm) \sin(2\tau_\pm+\phi_{\Gamma_\pm}), \\
&a_\pm^2-b_\pm^2=[\cos(2\tau_\pm+\phi_{\Gamma_\pm})-i \sin(2\tau_\pm+\phi_{\Gamma_\pm}) \sech(2\tau_\pm)]*
\exp\{i\phi_{\Gamma_\pm}-2i\arctan[\tanh(\tau_\pm)]\}.
\end{aligned}
\end{equation}
In the second case we have instead
\begin{equation} \label{Quantities def ro 2 case}
\begin{aligned}
&\text{Re}[a_+ b_+^*]={1 \over 2} \tanh(2\tau_+) \sin(2\tau_+ + \phi_{\Gamma_+}), \\
&a_+^2-b_+^2=[\cos(2\tau_+ + \phi_{\Gamma_+})-i \sin(2\tau_+ + \phi_{\Gamma_+}) \sech(2\tau_+)]*
\exp\{i\phi_{\Gamma_+}-2i\arctan[\tanh(\tau_+)]\}; \\
&\text{Re}[a_- b_-^*]={\tanh(\tau_-) \over \cosh(\tau_-)}\sin[\phi_{\Gamma_-}+\sinh(\tau_-)], \\
&a_-^2-b_-^2=[\sech^2(\tau_-)+e^{2i[\phi_{\Gamma_-}+\sinh(\tau_-)]}\tanh^2(\tau_-)]*\exp\{ -2i\arctan[\tanh(\tau_-/2)]-i\sinh(\tau_-) \}.
\end{aligned}
\end{equation}

We may consider now mixed states consisting of classical mixture of two Bell states, for example
\begin{equation}
\rho_0=p \ket{\Phi^+}\bra{\Phi^+} + (1-p) \ket{\Psi^\pm}\bra{\Psi^\pm}=p\rho_0^++(1-p)\tilde{\rho}_0^+={1 \over 2}
\left(
\begin{array}{cccc}
 p & 0 & 0 & p \\
 0 & 1-p & \pm(1-p) & 0 \\
 0 & \pm(1-p) & 1-p & 0 \\
 p & 0 & 0 & p \\
\end{array}
\right),\label{mixt1}
\end{equation}
with $0 \leq p \leq 1$.
Since the two subdynamics does not interfere it is easy to write the time evolution expression of the entries of $\rho(t)=U(t)\rho_0 U^\dagger(t)$, namely
\begin{equation}
\begin{aligned}
&\rho_{11}(t)=p\rho_{11}^+(t)= p\left({1 \over 2}+\text{Re}[a_+b_+^*]\right) , \quad
\rho_{14}(t)=p\rho_{14}^+(t)= p\left({ a_+^2-b_+^2 \over 2 }\right), \quad
\rho_{44}(t)=p\rho_{44}^+(t)= p\left({1 \over 2}-\text{Re}[a_+b_+^*]\right); \\
&\rho_{22}(t)=(1-p)\tilde{\rho}_{22}^+(t)= (1-p)\left({1 \over 2}\pm\text{Re}[a_-b_-^*]\right), \quad
\rho_{23}(t)=\pm(1-p)\tilde{\rho}_{23}^+(t)= (1-p)\left({ a_-^2-b_-^2 \over 2 }\right), \\
&\rho_{33}(t)=(1-p)\tilde{\rho}_{33}^+(t)= (1-p)\left({1 \over 2}\mp\text{Re}[a_-b_-^*]\right).
\end{aligned}\label{romixt1}
\end{equation}
Of course, the quantities $\text{Re}[a_\pm b_\pm^*]$ and $a_\pm^2-b_\pm^2$ defining the time-dependence of the entries in the two specific exactly solvable time-dependent scenarios are the same as those written in Eqs. \eqref{Quantities def ro 1 case} and \eqref{Quantities def ro 2 case}.
The analogous reasoning may be done for the other following two mixtures
\begin{equation}\label{mixt3}
\rho_0=p \ket{\Phi^-}\bra{\Phi^-} + (1-p) \ket{\Psi^\pm}\bra{\Psi^\pm}=p\rho_0^- + (1-p)\tilde{\rho}_0^\pm={1 \over 2}
\left(
\begin{array}{cccc}
 p & 0 & 0 & -p \\
 0 & 1-p & \pm(1-p) & 0 \\
 0 & \pm(1-p) & 1-p & 0 \\
 -p & 0 & 0 & p \\
\end{array}
\right),
\end{equation}
with
\begin{equation}
\begin{aligned}
&\rho_{11}(t)=p\rho_{11}^-(t)= p\left({1 \over 2}-\text{Re}[a_+b_+^*]\right) , \quad
\rho_{14}(t)=p\rho_{14}^-(t)= -p\left({ a_+^2-b_+^2 \over 2 }\right), \quad
\rho_{44}(t)=p\rho_{44}^-(t)= p\left({1 \over 2}+\text{Re}[a_+b_+^*]\right); \\
&\rho_{22}(t)=(1-p)\tilde{\rho}_{22}^\pm(t)= (1-p)\left({1 \over 2}\pm\text{Re}[a_-b_-^*]\right), \quad
\rho_{23}(t)=(1-p)\tilde{\rho}_{23}^\pm(t)= \pm(1-p)\left({ a_-^2-b_-^2 \over 2 }\right), \\
&\rho_{33}(t)=(1-p)\tilde{\rho}_{33}^\pm(t)= (1-p)\left({1 \over 2}\mp\text{Re}[a_-b_-^*]\right).
\end{aligned}
\end{equation}

The other two possible classical mixtures of Bell states are
\begin{equation}\label{mixt5}
\rho_0=p \ket{\Phi^+}\bra{\Phi^+} + (1-p) \ket{\Phi^-}\bra{\Phi^-}=p\rho_0^+ + (1-p)\rho_0^-={1 \over 2}
\left(
\begin{array}{cccc}
 1 & 0 & 0 & 2p-1 \\
 0 & 0 & 0 & 0 \\
 0 & 0 & 0 & 0 \\
 2p-1 & 0 & 0 & 1 \\
\end{array}
\right),
\end{equation}
and
\begin{equation}\label{mixt6}
\rho_0=p \ket{\Psi^+}\bra{\Psi^+} + (1-p) \ket{\Psi^-}\bra{\Psi^-}=p\tilde{\rho}_0^+ + (1-p)\tilde{\rho}_0^-={1 \over 2}
\left(
\begin{array}{cccc}
 0 & 0 & 0 & 0 \\
 0 & 1 & 2p-1 & 0 \\
 0 & 2p-1 & 1 & 0 \\
 0 & 0 & 0 & 0 \\
\end{array}
\right).
\end{equation}
In the first case we have
\begin{equation}
\begin{aligned}
\rho_{11}(t)= {1 \over 2}+(2p-1)\text{Re}[a_+b_+^*] , \quad
\rho_{14}(t)= { (2p-1) \over 2 }(a_+^2-b_+^2), \quad
\rho_{44}(t)= {1 \over 2}-(2p-1)\text{Re}[a_+b_+^*],
\end{aligned}\label{romixt5}
\end{equation}
while in the second one we get
\begin{equation}\label{romixt6}
\begin{aligned}
\rho_{22}(t)= {1 \over 2}+(2p-1)\text{Re}[a_-b_-^*], \quad
\rho_{23}(t)= { (2p-1) \over 2 }(a_-^2-b_-^2), \quad
\rho_{33}(t)= {1 \over 2}-(2p-1)\text{Re}[a_-b_-^*].
\end{aligned}
\end{equation}

\end{document}